\documentclass[a4paper,11pt]{article}
\usepackage{jcappub}
\usepackage{xcolor}
\usepackage[normalem]{ulem}
\usepackage{enumitem}
\usepackage{kantlipsum}
\usepackage{mathtools}

\makeatletter
\newcommand{\preprint}[1]{\gdef\@preprint{#1}}
\preprint{} % Initialize as empty
\patchcmd{\maketitle}{\@fpheader}{\@fpheader\hfill\@preprint}{}{}
\makeatother
\preprint{ULB-TH/26-03}

\usepackage{amsmath}	% AMS Math Package
\usepackage{amsthm}		% Theorem Formatting
\usepackage{amssymb}	% Math symbols such as \mathbb
\usepackage{array}
\usepackage{datetime}
\usepackage{graphicx}
\usepackage{color}
\usepackage{verbatim}
\usepackage{bm}
\usepackage{braket}
\usepackage{natbib}
\usepackage{cancel}
\usepackage{xcolor}
\usepackage{slashed}
\usepackage{braket}
\usepackage{amsfonts} 
\smallskipamount
\usepackage{feynmp}
\usepackage{letltxmacro} % closed roots
\usepackage{setspace}

\usepackage{tikz,xcolor}
\definecolor{lime}{HTML}{A6CE39}
\DeclareRobustCommand{\orcidicon}{
	\begin{tikzpicture}
	\draw[lime, fill=lime] (0,0) 
	circle [radius=0.16] 
	node[white] {{\fontfamily{qag}\selectfont \tiny ID}};
	\draw[white, fill=white] (-0.0625,0.095) 
	circle [radius=0.007];
	\end{tikzpicture}
	\hspace{-2mm}
}
\foreach \x in {A, ..., Z}{
	\expandafter\xdef\csname orcid\x\endcsname{\noexpand\href{https://orcid.org/\csname orcidauthor\x\endcsname}{\noexpand\orcidicon}}
}

\settimeformat{ampmtime}

\definecolor{grey}{rgb}{0.4,0.4,0.4}
\definecolor{dullmagenta}{rgb}{0.4,0,0.4}
\definecolor{darkblue}{rgb}{0,0,0.4}
\definecolor{midblue}{rgb}{0,0,0.5}
\definecolor{midred}{rgb}{0.5,0,0}
\definecolor{orange}{rgb}{1,0.5,0}
\definecolor{lightbrown}{rgb}{0.75,0.5,0.25}
\definecolor{tan}{cmyk}{0.14,0.42,0.56,0}
\definecolor{djunglegreen}{cmyk}{0.99,0,0.52,0}
\definecolor{lightgreen}{rgb}{0,1,0}
\definecolor{olivegreen}{cmyk}{0.64,0,0.95,0.40}
\definecolor{midgreen}{rgb}{0.0,0.675,0.0}
\definecolor{darkgreen}{rgb}{0,0.5,0}

\newcommand{\vs}{\vspace}

\renewcommand{\.}{\hspace{0.5mm}}

\newcommand{\Frm}{\ensuremath{\mathrm{F}}}

\newcommand{\grm}{\ensuremath{\mathrm{g}}}

\newcommand{\srm}{\ensuremath{\mathrm{s}}}

\renewcommand{\d}{\ensuremath{\mathrm{d}}}
\settimeformat{ampmtime}

\makeatletter
\let\oldr@@t\r@@t
\def\r@@t#1#2{%
\setbox0=\hbox{$\oldr@@t#1{#2\,}$}\dimen0=\ht0
\advance\dimen0-0.2\ht0
\setbox2=\hbox{\vrule height\ht0 depth -\dimen0}%
{\box0\lower0.4pt\box2}}
\LetLtxMacro{\oldsqrt}{\sqrt}
\renewcommand*{\sqrt}[2][\ ]{\oldsqrt[#1]{#2}}
\makeatother

\newcommand{\Ts}{{T_\star}}
\newcommand{\e}{{\rm ev}}
\newcommand{\dm}{{\rm DM}}
\newcommand{\bh}{{\rm BH}}

\title{Non-Cold Dark Matter from Memory-Burdened Primordial Black Holes}

\author[a,b,c]{Valentin Thoss,}
\affiliation[a]{
Universit\"ats-Sternwarte,
Ludwig-Maximilians-Universit\"at M\"unchen,
Scheinerstr. 1,
81679 Munich, Germany}
\affiliation[b]{
Max-Planck-Institut f\"ur Extraterrestrische Physik,
Gießenbachstraße 1,
85748 Garching, Germany}
\affiliation[c]{
Excellence Cluster ORIGINS,
Boltzmannstraße 2,
85748 Garching, Germany
}

\author[d, e]{Laura Lopez-Honorez,}

\affiliation[d]{Service de Physique Th\'{e}orique \& Brussels Laboratory of the Universe BLU-ULB, CP 225, Universit\'{e} Libre de Bruxelles,  Boulevard du Triomphe, B-1050 Brussels, Belgium}

\affiliation[e]{Theoretische Natuurkunde \& The International Solvay Institutes, \\
Vrije Universiteit Brussel, Pleinlaan 2, B-1050 Brussels, Belgium}

\author[f,g,h]{Florian K{\"u}hnel,}
\affiliation[f]{
	Max-Planck-Institut f{\"u}r Physik,
	Boltzmannstr.~8,
	85748 Garching,
	Germany}
\affiliation[g]{
	Arnold Sommerfeld Center,
	Ludwig-Maximilians-Universit{\"a}t,
	Theresienstr.~37,
	80333 M{\"u}nchen,
	Germany}
\affiliation[h]{
	Fakult{\"a}t Physik,
	Technische Universit{\"a}t Dortmund,
	August-Schmidt-Str.~4,
	44221 Dortmund,
	Germany}

\author[d]{and Marco Hufnagel}

\emailAdd{vthoss@mpe.mpg.de, llopezho@ulb.be, fkuehnel@mpp.mpg.de, marco.hufnagel@ulb.be}

\abstract{
Non‑cold dark matter particles can arise from the evaporation of primordial black holes (PBHs). In this paper, we further investigate how the memory‑burden effect, which delays the full evaporation of black holes, affects the Lyman-$\alpha$ bound on such non‑cold dark matter (NCDM) particles. We mainly focus on scenarios in which PBHs have fully evaporated by today, undergoing a semi‑classical evaporation phase followed by a memory‑burden dominated phase. In this framework, PBH evaporation generically leads to two distinct dark matter populations with different velocity dispersions, which can imprint observable signatures on the matter power spectrum. We compute the resulting NCDM phase‑space distribution and its impact on small‑scale overdensities using the {\tt BlackHawk} and {\tt CLASS} codes. This is then used to reinterpret Lyman‑$\alpha$ forest constraints for thermal warm dark matter, deriving both a velocity‑dispersion‑based and a matter‑power‑spectrum‑based estimate. In particular, we discuss how we obtain constraints on scenarios in which NCDM particles constitute only a fraction of the total relic dark matter. Finally, we discuss the viable parameter space as a function of dark matter masses, PBH initial conditions, and memory‑burden parameters. We show that even subdominant NCDM components from PBH evaporation can be constrained, and confirm that NCDM can only account for all of the dark matter in the absence of PBH domination, as in the semi‑classical case.
}

\begin{document}

\maketitle
\flushbottom

%%%%%%%%%%%%%%%%%%%%%%%%%%%%%%%%%%%%%%%
\section{Introduction}
\label{sec:intro}
%%%%%%%%%%%%%%%%%%%%%%%%%%%%%%%%%%%%%%%

Since their proposal about half a century ago~\cite{Hawking:1971ei, Carr:1974nx}, primordial black holes (PBHs) have been subject to a long series of investigations, in particular in their role as a potential dark matter (DM) candidate (see Refs.~\cite{Carr:2016drx, Carr:2020xqk, Escriva:2022duf} for reviews). This concerns their formation mechanisms as well as the signatures they leave, ranging from gravitational lensing and dynamical effects to gravitational waves and evaporation. The latter, which is particularly relevant for low masses and is thought to observationally exclude PBHs from constituting a significant DM fraction (cf.~Ref.~\cite{Carr:2016hva}), has recently been revisited (see, e.g.,~Refs.~\cite{Alexandre:2024nuo, Thoss:2024hsr}). According to those analyses, black holes (BHs) do not evaporate semi-classically (SC) all the way until or near complete evaporation, but show a dramatic slow-down of their evaporation rate at the latest after half of their mass has been radiated away{\,---\,}possibly much before. The underlying reason is the recently-proposed {\it memory-burden} (MB) effect~\cite{Dvali:2018xpy, Dvali:2020wft}, which has become a subject of intense studies (see, e.g., Refs.~\cite{Alexandre:2024nuo, Thoss:2024hsr, Dvali:2024hsb, Bhaumik:2024qzd, Kohri:2024qpd, Chianese:2024rsn, Zantedeschi:2024ram, Chaudhuri:2025asm, Montefalcone:2025akm, Dvali:2025ktz}), indicating the possibility for a new window of ultra-light PBH DM.\footnote{Besides the mentioned (standard) MB effect, Dvali~\cite{Dvali:2025sog} has recently pointed out the existence of a "swift" pendant, which concerns BH perturbations, and should leave observable signatures.}

Regardless of whether PBHs are such DM (quasi-)relics, any modification of the BH evaporation could potentially have far-reaching consequences, as the high temperatures of ultra-light PBHs could have led to the injection of various DM particles with the potential to interfere with 21-cm physics~(cf.~Ref.~\cite{Tashiro:2012qe, Clark:2018ghm, Sun:2025ksr}), the Lyman-$\alpha$ forest~\cite{Murgia:2019duy, Ivanov:2025pbu}, and/or Big Bang Nucleosynthesis (BBN)~\cite{Miyama:1978mp, Keith:2020jww,Wang:2025pum}. Particularly interesting imprints on the latter two are through a population of non-cold DM (NCDM) from evaporating PBHs of sufficiently low mass, see, e.g., Refs.~\cite{Fujita:2014hha, Allahverdi:2017sks, Lennon:2017tqq, Masina:2020xhk, Baldes:2020nuv, Auffinger:2020afu, Barman:2024iht} for  discussions.

Those effects clearly depend on the BH evaporation dynamics. In particular, the knowledge of when and how the transition to the MB phase sets in is fundamental in quantifying the mentioned imprint of ultra-light PBHs. While most studies have assumed an instant MB onset at an order-one fraction of mass-loss, it has now become clear that the transition is a) smooth and b) might set in already on a much shorter time scale, related to inverse powers of the BH entropy (see Refs.~\cite{Dvali:2020wft, Dvali:2024hsb, Dvali:2025ktz}).

Various studies have explored DM production from PBH evaporation, see, e.g., Refs.~\cite{Matsas:1998zm, Dolgov:2000ht, Bell:1998jk, Baumann:2007yr, Dai:2009hx, Fujita:2014hha, Morrison:2018xla, Green:1999yh, Allahverdi:2017sks, Hooper:2019gtx, Gondolo:2020uqv, Masina:2020xhk, Auffinger:2020afu, Bernal:2020bjf, Khlopov:2004tn, Chaudhuri:2020wjo, Baldes:2020nuv, Bernal:2020ili, Bernal:2020kse, Bernal:2021bbv, Samanta:2021mdm, Cheek:2021cfe, Cheek:2021odj, Barman:2021ost, Chen:2023tzd,Bertuzzo:2024fns,Shallue:2024hqe}. More recently, the impact of MB on the latter has been investigated in e.g.~Refs.~\cite{Haque:2024eyh,Barman:2024iht,Kitabayashi:2025iaq,Borah:2024bcr}, studying among others the possibility of producing DM alongside the baryon asymmetry. No detailed analysis of the NCDM velocity distribution and associated constraints has yet been provided. Interestingly, in Ref.~\cite{Barman:2024iht}, it was argued that the simultaneous production of both DM and the baryon asymmetry is not feasible due to Lyman-$\alpha$ constraints associated with the NCDM imprint. Yet, their evaluation of such a constraint is based on a rough estimation of the typical velocity of DM particles based on a previous simple estimate of Ref.~\cite{Masina:2020xhk} that we revise here. 

In this paper, we provide a more detailed investigation of Lyman-$\alpha$ constraint following a methodology similar to~\cite{Lennon:2017tqq, Baldes:2020nuv}. For that purpose, we evaluate the phase-space distribution of NCDM arising from PBH evaporation using {\tt BlackHawk}~\cite{Arbey:2019mbc} to obtain the correct spectrum of evaporated particles. The latter is used as input for the Boltzmann code {\tt CLASS}~\cite{Lesgourgues:2011rh}, which then allows to extract the matter power spectrum, and to analyse its deviation with respect to $\Lambda$CDM. In turn, this enables us to constrain the DM from PBH evaporation using the structure-formation bounds from Lyman-$\alpha$ data.

This article is structured as follows: After the introduction in Sec.~\ref{sec:intro}, Sec.~\ref{sec:evap-burden} discusses aspects of BH evaporation, including the MB effect. Section~\ref{sec:DMabund} elaborates on PBH abundance limits from the DM abundance in general, while Sec.~\ref{sec:DMevap} investigates the implications of the generation of non-cold particle DM from PBH evaporation, and Sec.~\ref{sec:param-space} is dedicated to determining the allowed parameter space. Finally, we conclude in Sec.~\ref{sec:concl}.

%%%%%%%%%%%%%%%%%%%%%%%%%%%%%%%%%%%%%%%
\section{Black Hole Evaporation and Memory Burden}
\label{sec:evap-burden}
%%%%%%%%%%%%%%%%%%%%%%%%%%%%%%%%%%%%%%%

In this Section we review the physics of BH evaporation and how it is modified by the MB effect. In the following, $M_{\Frm}$ and $T_{\Frm}$ denote the  mass and temperature of the BH at formation, which for non-rotating BHs are related via the relation
\begin{equation}
    T_{\Frm}
        =
            \frac{M_{\rm P}^{2}}{8\pi M_{\Frm}}\,,
    \label{eq:TBH}
\end{equation}
where $M_{\rm P} = 1.22\times 10^{19}\,$~GeV is the Planck mass. On the other hand, the Bekenstein--Hawking entropy of a BH, $S$, is related to the BH mass, $M$, through 
\begin{equation}
    S
        =
            4\pi\,\frac{M^{2}}{M_{\rm P}^{2}}\,.
    \label{eq:S}
\end{equation}

%%%%%%%%%%%%%%%%%%%%%%%%%%%%%%%%%%%%%%%
\subsection{Semi-Classical Evaporation}
\label{sec:SC}
%%%%%%%%%%%%%%%%%%%%%%%%%%%%%%%%%%%%%%%

The SC emission rate of a neutral and non-rotating BH with mass $M$ is given by~\cite{Hawking1975}
\begin{equation}
    \frac{\mathrm{d}N_j^{\rm sc}}{\mathrm{d}E\.\mathrm{d}t}
        =
            \frac{g_j}{2\pi}\frac{\Gamma(E,M,s_j)}{e^{E/T} - (-1)^{2s_j}}\,,
    \label{eq:dNdEdt_SC}
\end{equation}
where  $\,g_j$ and $s_j$ denote the number of degrees of freedom and the spin of the particle species $j$ emitted from the BH, and $\Gamma$ encodes the Greybody factors. In the limit $E\gg T$, known as the geometric-optics limit, they approach $\Gamma = 27E^{2}M^{2}/M_{\rm P}^{4}$ but otherwise depend on $E/T$ and the spin $s_j$ of the emitted particle. Note that for the SC phase, the temperature appearing in the exponential factor is always assumed to be the time-dependent BH temperature, which scales as the inverse of the PBH mass, $\propto M( t )^{-1}$. The temperature thus increases during the evaporation process, which{\,--- \,}in particular{\,--- \,}leads to a high-energy tail in the DM-momentum distribution in the final stages of BH evaporation~\cite{Lennon:2017tqq}.

In this work, we use the public code {\tt BlackHawk}~\cite{Arbey:2019mbc} to accurately evaluate the BH evaporation spectra. In particular, the BH mass decreases with time at a rate
\begin{equation}
    \frac {\d M}{\d t }
        =
            -\sum_j\int_{0}^{\infty} E \frac {\d N_j^{\rm sc}}{\d t\.\d E}\;
    \d E
        =
            - e_T\frac{M_{\rm P}^{4}}{M^{2}}\,,
        \label{eq:dMdt}
\end{equation}
with 
\begin{equation}
    e_T
        =
            4.3\times 10^{-3}\qquad {\rm for}\qquad T_{\rm BH} >T_{\rm EW} \,, 
  \label{eq:eT}
\end{equation}
where the prefactor $e_T$ accounts for all the evaporated degrees of freedom and has been obtained with {\tt BlackHawk}. For definiteness, we have assumed here that the BH  emits a two-component fermionic DM particle with mass $m_{\rm DM} < T_{\rm BH}$ together with all Standard Model (SM) particles and that the temperature of the BH is above the electroweak scale, $T_{\rm BH} > T_{\rm EW}$. In the following, unless otherwise stated, we will assume that $T_{\rm BH} >T_{\rm EW}$, which translates to $M_{\Frm} \lesssim 10^{10}$\,g, and that our DM particle is a fermionic species with two degrees of freedom, i.e. $g_{\rm DM}=2$, in all relevant numerical results or illustrative plots.

From Eq.~(\ref{eq:dMdt}), the BH mass evolves with time as follows
\begin{equation}
    M( t )
        = 
            M_{\Frm}\left( 1- \frac{(t-t_{\Frm})}{t_{\rm sc}} \right)^{1/3}\,,
  \label{eq:MBH_SC}
\end{equation}
with $t_{\Frm}$ being the time of formation. The BH lifetime, $t_{\rm sc}$, reads
\begin{eqnarray}
    t_{\rm sc}
        &=&
            \frac{1}{3\.e_T} \frac{M_{\Frm}^{3}}{M_{\rm P}^{4}}\simeq 4.2 \times 10^{-12}\,\srm \left(\frac{M_{\Frm}}{10^{10}M_{\rm P}}\right)^{3}\,.
  \label{eq:tau}
\end{eqnarray}
Here we see that the BH lifetime is fixed by its initial mass $M_{\Frm}$. Furthermore, by integrating Eqs.~(\ref{eq:dNdEdt_SC}) 
over time and energy, we can compute the total number of particles $j$ 
emitted during the PBH lifetime,  $N^{\rm sc}_j$. The latter is given by
\begin{eqnarray}
    N^{\rm sc}_j
        &=&
            2.8\times 10^{-2}\,g_j \frac{M_{\Frm}^{2}}{M_{\rm P}^{2}}  \qquad \text{for} \qquad T_\bh>T_{\rm  EW} \,, t\geq t_\e\, ,\nonumber
\end{eqnarray}
for a particle of spin 1/2. Note that the number of particles emitted from the BH thus simply scales as $M_{\rm P}^{2}/T_{\Frm}^{2}$.

%%%%%%%%%%%%%%%%%%%%%%%%%%%%%%%%%%%%%%%
\subsection{The Memory-Burden Effect}
\label{sec:MB}
%%%%%%%%%%%%%%%%%%%%%%%%%%%%%%%%%%%%%%%

The physics studied in this paper depends fundamentally on the BH evaporation dynamics, which so far, and in the vast amount of literature, has been treated in a SC way. However, this regime is certainly invalidated at some point during evaporation. Under no circumstances could the SC regime be valid until the BH mass becomes Planckian. However, as discussed in Refs.~\cite{Dvali:2018xpy, Dvali:2020wft}, this should happen already much earlier. Since the associated scale has not yet been determined, in this work, we will remain general and denote the mass scale at which the SC approach is no longer valid as $q\.M_{\Frm}$. However, note that in Refs.~\cite{Dvali:2018xpy, Dvali:2020wft}, it was argued that the MB should set in at the latest when the BH has lost half of its mass, i.e.~for $ q \geq 0.5$. 

The {\it memory-burden} effect is actually a property of generic systems with high microstate entropy $S$, which have an exponentially large microstate degeneracy due to nearly-gapless "memory" modes. Various excitation patterns of these encode the information carried by the system; the high information capacity is due to the low energy cost and the large number of memory modes. Decay of the system lifts the mode-degeneracy, which leads to an increase of the gap and therefore of the cost to store information, thereby slowing down the decay. This is the essence of the MB effect (see Refs.~\cite{Dvali:2018xpy, Dvali:2020wft}).

For BHs, the inevitability of this effect is particularly clear when starting with the (wrong) assumption that their evaporation is self-similar. This would imply that after the BH lost half of its mass, their radius must have shrunken by $50\%$ and its entropy by $75\%$ (because it is proportional to the surface area). In the SC limit, the emitted radiation has a thermal spectrum, which{\,---\,}by its very nature{\,---\,}cannot possibly account for the enormous loss of information implied by the decrease in entropy. This undermines the assumption that the decay process is self‑similar.
\begin{comment}
    \llh{This is not as sentence:} Now, in the SC limit, a thermal emission spectrum, which {\,---\,}{\it under no circumstances}{\,---\,} could explain the huge loss of information capacity accompanying the decrease in entropy. This invalidates the assumption of self-similarity of the decay.
\end{comment}

Reference~\cite{Dvali:2020wft} has both analytically and numerically studied the MB effect, and argued for a drastic drop of the evaporation rate, which becomes entropy-suppressed
\begin{equation}
    \frac{\mathrm{d}N^{\rm mb}}{\mathrm{d}E\.\mathrm{d}t}
        =
            \frac{1}{S^k}\frac{\mathrm{d}N^{\rm sc}}{\mathrm{d}E\.\mathrm{d}t}\, ,
     \label{eq:Gamma}
\end{equation}
with the subscripts 'sc' and 'mb' indicating the semi-classical and memory-burden phases. Currently, the exact value of the exponent $k$ is undetermined, with Ref.~\cite{Dvali:2020wft} suggesting $1 \lesssim k \lesssim 3$. In this paper, we will remain general, considering $k\geq 0$. Note that the above suppression is enormous: $S^{-k} \sim 10^{-30\.k}\,( M / 10^{10}\.\grm\. )^{-2\.k}$. This increases the BH lifetime by many orders of magnitude, e.g.~by a factor $\sim 10^{30}$ even for $k = 1$ and $M =  10^{10}\.\grm$, which in turn could open up a large window for light PBH DM (see, e.g., Refs.~\cite{Thoss:2024hsr, Alexandre:2024nuo, Dvali:2024hsb}.)

%%%%%%%%%%%%%%%%%%%%%%%%%%%%%%%%%%%%%%%
\subsection{Semi-Classical Evaporation Followed by a Memory-Burden Stage}
\label{sec:SC+MB}
%%%%%%%%%%%%%%%%%%%%%%%%%%%%%%%%%%%%%%%

Using the SC equations, one can show that the time at which the MB effect sets in, i.e.~the time at which $M = q M_{\Frm}$, is given by $t_{\Frm}+t_{q}$ with
\begin{equation}
    t_{q}
        =
            (1-q^{3})\,t_{\rm sc}\, .
  \label{eq:tq}
\end{equation}
During the SC phase, i.e.~prior to the onset of MB effects, the BH will have produced a number
of particles $N_{q}$ with
\begin{equation}
    N_{q}
        =
            (1-q^{2})\,N_{\rm sc}\,,
  \label{eq:Nq}
\end{equation}
where we have omitted the index $j$ for the particle species $j$ for simplicity.

As discussed above, during the MB stage of evaporation, the evaporation rate for a species $j$ is expected to be suppressed by $k$ powers of the entropy. However, at this point, there is no clear indication how the relation between the BH mass and the energy of the emitted quanta evolves when the SC description breaks down, see e.g.~the discussion in Ref.~\cite{Thoss:2024hsr}. In this paper, we consider two prescriptions for the evaporation rate during the MB phase. In most cases, we will assume that the temperature of the BH remains fixed as $T = T_{\Frm}/q$ during the MB phase with 
\begin{equation}
    \frac{\mathrm{d}N^{\rm mb}_{ j}}{\mathrm{d}E\.\mathrm{d}t}
        =
            \frac{g_j}{2\pi}\frac{1}{S(q\.M_{\Frm})^{k}} \frac{\Gamma(E,q\.M_{\Frm},s_j)}{e^{qE/T_\Frm} - (-1)^{2s_j}}\,, \quad[{\rm no \,burst}]
    \label{eq:dNdEdt_MB}
\end{equation}
where an extra factor of $k$ powers of the entropy density $S(q M_\text{F})$ at fixed mass $q\.M_{\Frm}$ has been introduced in the denominator and the argument of the exponential factor is set to $q\.E/T_\Frm$. This, in particular, suppresses the high-energy tail in the momentum distribution of particles evaporating from the BH. We will also discuss the possibility that the energy of the quanta emitted in the MB phase continues to increase as the BH evaporates, as in the SC approach. In the following, we will refer to this scenario as evaporation with burst, with an emission rate given by
\begin{equation}
    \frac{\mathrm{d}N^{\rm mb}_{ j}}{\mathrm{d}E\.\mathrm{d}t}
        = 
            \frac{g_j}{2\pi}\frac{1}{S(M)^{k}} \frac{\Gamma(E,M,s_j)}{e^{E/T} - (-1)^{2s_j}}\,.\quad[{\rm burst}]
    \label{eq:dNdEdt_MB_burst}
\end{equation}
We again have an extra factor of $k$ powers of the entropy density in the denominator, but this time it is $S = S(M)$, which depends on the time-dependent BH mass, $M(t)$. The argument of the exponential factor also scales as $E/T$, where $T$ denotes the BH temperature scaling as $M^{-1}$.

\begin{figure}[t]
    \centering
    \includegraphics[width = 0.7\linewidth]{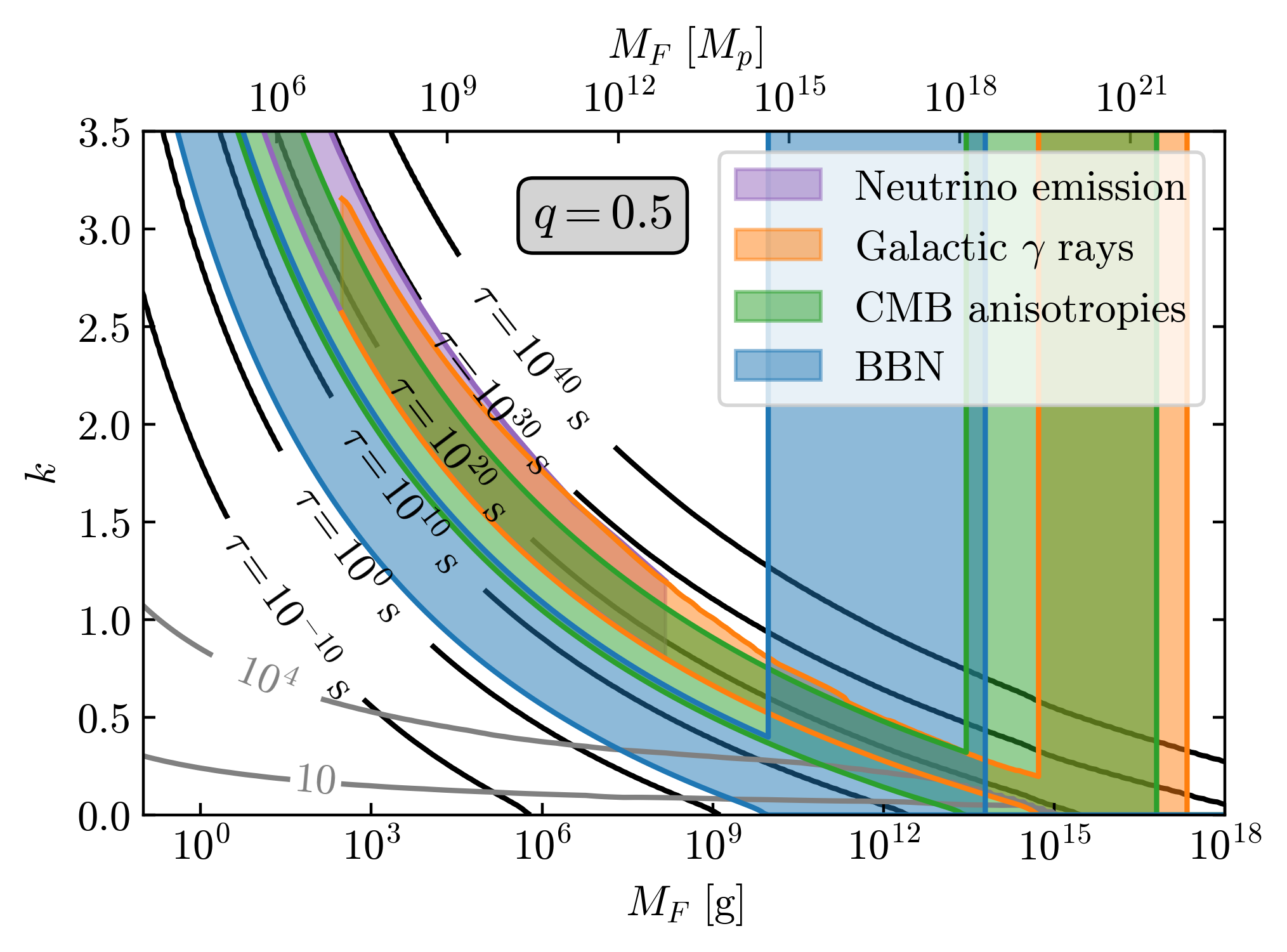}
   \caption{ Case of BH evaporation with $q = 0.5$ assuming an instantaneous transition ($\delta = 0$) without burst. Black contours represent constant lifetimes $\tau = t_{\rm mb}$ for PBHs subject to the MB effect. Gray contours indicate ratios of scale factors $a_{\rm mb}/a_{\rm q}$ of 10 and $10^{4}$, with $a_{\rm mb}$ the scale factor at full evaporation considering MB and $a_{\rm q}$ the one at the beginning of the MB phase. For low values of this ratio, one would expect comparable root-mean-squared velocities for particles emitted both in the MB and SC phases. The color bands show constraints from neutrino emissions, galactic gamma rays, CMB anisotropies and BBN as computed in Refs.~\cite{Thoss:2024hsr,Chaudhuri:2025asm}. See text for more details.
 }
    \label{fig:lifetimes}
\end{figure}

During the MB phase without burst, using Eq.~\eqref{eq:dNdEdt_MB} we can show that the emission rate is constant and that the BH mass evolves as
\begin{equation}
    M( t )
        = 
            q\,M_{\Frm} \left(1- \frac{(t-t_{q})}{t_{\rm mb}}\right)\qquad {\rm for }\qquad t>t_{q}\,, \quad[{\rm no \,burst}]
    \label{eq:MBH_MB}
\end{equation}
where $t_{\rm mb}$ is the BH lifetime in the MB phase. In contrast, in the case of a BH burst, one can show that the BH mass evolves as
\begin{equation}
    M( t )
        = 
            q\,M_{\Frm} \left(1- \frac{(t-t_{q})}{t_{\rm mb}}\right)^{\!1/(3+2k)}\qquad {\rm for }\qquad t>t_{q}\,. \quad[{\rm burst}]
    \label{eq:MBH_MB_burst}
\end{equation}
The BH lifetime in the MB phase reads
\begin{eqnarray}
    t_{\rm mb}
        &=&
            \frac{3\,q^{3}\.S(q\.M_{\Frm})^{k}}{\kappa}\; t_{\rm sc}\,.
    \label{eq:tautild}
\end{eqnarray}
where 
\begin{equation}
    \kappa
        =
            \begin{dcases*}
                1   & [no burst]\,,\\[1mm]
                3+2k&  [burst]\,.
            \end{dcases*}
    \label{eq:kappa}
\end{equation}
Note that in the limit $q \to 0$ one recovers the SC case. Increasing $k$ can significantly increase the total lifetime of the BH. This can be seen in Fig.~\ref{fig:lifetimes} where contours of constant lifetime are shown with black continuous lines in the $(M_{\Frm},k)$ plane. In particular, in the case of $q = 0.5$, $k = 2$ without burst, one finds that BHs with $M_{\rm F}=2.2 \times 10^{8}M_{\rm P}$ would have evaporated approximately today with
\begin{equation}
    t_{q = 0.5}
        \simeq
            3.9\times 10^{-17}\,\srm \left(\frac{M_{\Frm}}{2.2\times 10^{8}M_{\rm P}}\right)^{\!3} \quad {\rm and} \quad 
    t_{\rm mb}
        \simeq
            3.9 \times 10^{17}\,\srm \left(\frac{M_{\Frm}}{2.2\times 10^{8}M_{\rm P}}\right)^{\!7}\,.
\end{equation}

The number of particles emitted per mass loss of the BH scales with the BH temperature as $\frac{\mathrm{d}N_i}{\mathrm{d}M}\sim g_iT_{\rm BH}^{-1}$, independently of the suppression in the evaporation rate. Therefore, the total number of particles emitted during the MB phase depends only on whether the BH temperature stays constant (no burst) or increases (burst) during the evaporation. For a constant temperature (no burst), we have $N_i\propto \int \mathrm{d}M M_{\Frm} = M_{\Frm}^{2}$, that is a factor of two higher than in the case of a burst with $N_i\propto \int \mathrm{d}M M = M_{\Frm}^{2}/2$. Therefore, the number of particles emitted during the MB phase is given by 
\begin{equation}
    N_{\rm mb}
        =
            \xi\.q^{2}N_{\rm sc}\,,
\end{equation}
with 
\begin{equation}
	\xi
        =
            \begin{dcases*}
                2   & [no burst]\,,\\[1mm]
                1&  [burst]\,.
            \end{dcases*}
    \label{eq:xi}
\end{equation}
In particular, this implies that in the case where MB effects set in when the PBH has reached $q M_{\rm F}$ and where the PBH has fully evaporated through a SC and a MB phase, we have a total number of evaporated particles equal to $(1-q^{2}) N_{\rm sc} + \xi q^{2} N_{\rm sc}$. This is larger than $N_{\rm sc}$ in the case without burst as $\xi = 2$ while it is equal to $N_{\rm sc}$ in the case with burst.

Also note that in most phenomenological studies to date, the transition from the SC to the MB phase has been approximated as being instantaneous. Recent works~\cite{Montefalcone:2025akm, Dvali:2025ktz} have investigated the impact of a smooth transition with width $\Delta M/M_{\Frm}\sim \delta$ on PBH abundance constraints, see Sec.~\ref{sec:obs-cons-PBH} below. It turns out that a non-zero $\delta$ can have a potentially strong effect on the observational constraints. The magnitude of $\delta$ and its relation to $q$ is still a matter of active research. It has been argued that
\begin{equation}
    \delta
        \sim
            \mathcal{O}(0.1) \times (1-q)
    \label{eq:transition}
\end{equation}
(for dimensional reasons and from advanced toy models) and that both $\delta$, and $1-q$, are proportional to inverse powers of the black hole entropy $S$, such as e.g.~$\delta\propto 1/\sqrt{S}$. Since the entropy is generally huge for BHs, $S \simeq 10^{30}\,( M / 10^{10}\,\grm )^{2}$, the transition might be rather rapid, reinvigorating the sudden approximation. In the limit $\delta\rightarrow 0$, we also have $q\rightarrow 1$, corresponding to a very early onset of MB.

Throughout most of this work, we focus on the scenario of fully evaporated PBHs. In this case, the nature of the transition from the SC to the MB phase has little consequence on our results. For the computation of the DM abundance (see Section~\ref{sec:DMabund}), the relevant physical parameters are the lifetime of the PBHs, denoted $t_{\rm ev}$ in general, and the number of DM particles produced by the evaporating PBHs, $N_{\rm DM}$. The width $\delta$ of the transition has only a very small effect on the overall lifetime of the PBH for $\delta\lesssim 0.1$, as it is dominated by the time spent in the MB phase after the transition. The number of emitted particles is independent of the strength of the suppression during the MB phase and the transition regime. It only depends on the evolution of the BH temperature (burst/no burst case) up to a factor two, see Eq.~(\ref{eq:xi}). In principle, the spectra of the evaporating BH, discussed in Section~\ref{sec:DMevap}, are modified in the case of a non-instantaneous transition. However, we have checked that the correction to the constraints from the Lyman-$\alpha$ forest are of the order of $\delta$ and thus typically small.\footnote{Note that in the exponential parametrization in Ref.~\cite{Dvali:2025ktz} $\delta$ can in principle be arbitrarily large but for $\delta \gg 1$ one simply recovers a constant evaporation rate throughout the black hole's lifetime, equal to the initial SC rate. In the $\tanh$ parametrization of Ref.~\cite{Montefalcone:2025akm} $\delta\gtrsim 1/q$ would lead to a strong suppression from the onset of evaporation, which is unphysical.} This is simply a result of the fact that these bounds are well determined by the mean and root-mean-squared momentum of the DM particles, which only change by a factor of order $1 - \delta$ if one accounts for the transition, see Sec.~\ref{sec:DMevap} for more details.

%%%%%%%%%%%%%%%%%%%%%%%
\subsection{Observational Constraints on Evaporating Black Holes}
\label{sec:obs-cons-PBH}
%%%%%%%%%%%%%%%%%%%

\begin{figure}
    \centering
    \includegraphics[width = 0.49\linewidth]{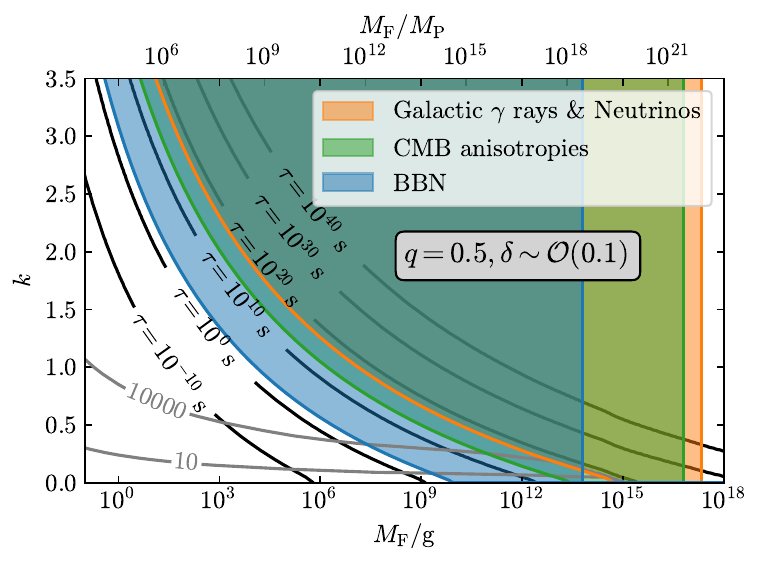}
    \includegraphics[width = 0.49\linewidth]{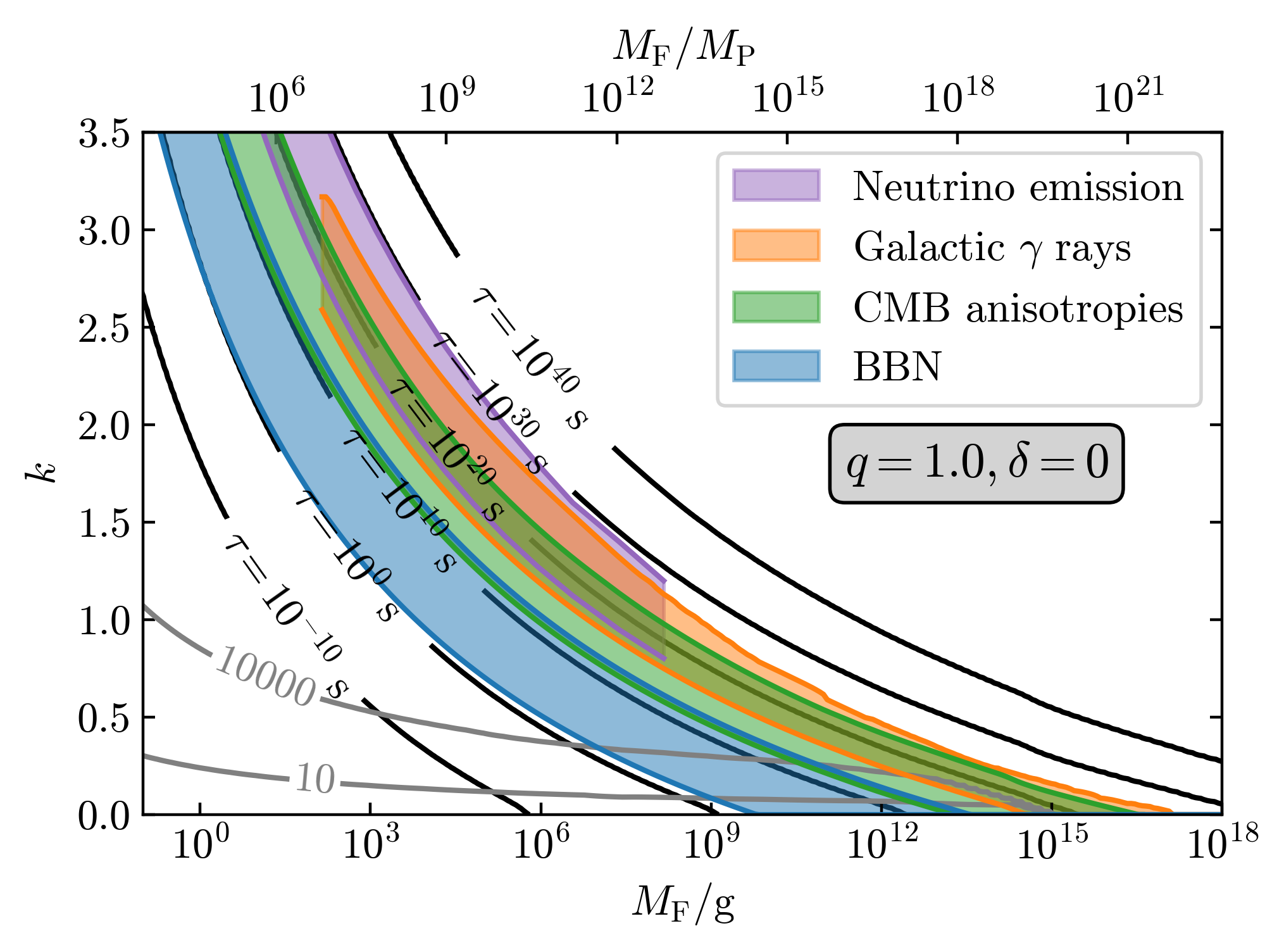}
    
    \caption{Illustration of the change in the constraints coverage of the parameter space when varying $q$, fixing the time at which MB sets in, and $\delta$, fixing the smoothness of the transition. Left: Case of  $q = 0.5$ assuming a smooth transition with e.g.~$\delta \sim \mathcal{O}(0.1)$ without burst. Right: Case of BH evaporation with $q = 1$ and $\delta = 0$. The colored lines and contours are equivalent to the ones of Fig.~\ref{fig:lifetimes}.}
    \label{fig:Mkplot-transitions}
\end{figure}

Based on results from \cite{Thoss:2024hsr, Chaudhuri:2025asm, Chianese:2024rsn}, Figs.~\ref{fig:lifetimes} and \ref{fig:Mkplot-transitions} provide an overview of observational constraints of evaporating PBHs that accounts for the MB effect assuming different $q$ and $\delta$ values.\footnote{Note that Ref.~\cite{Kim:2025kgu} recently obtained new constraints on PBHs fully evaporating before BBN due to their isocurvature perturbations contribution constrained by Planck data~\cite{Aghanim:2018eyx}. This analysis was, however, done in the case of SC evaporation and left the study of such constraints in the MB case for future work. We have thus not accounted for the latter here.} In particular, Refs.~\cite{Thoss:2024hsr, Chaudhuri:2025asm} computed constraints on the abundance of evaporating PBHs based on their effect on galactic $\gamma$-ray emissions, CMB anisotropies, and BBN observables. Reference~\cite{Chianese:2024rsn} extended this analysis by considering the emission of neutrinos from PBHs. 

Figure~\ref{fig:lifetimes} projects the exclusion regions from BBN (blue), CMB anisotropies (green), $\gamma$ rays (orange), and neutrino observations (purple) in the plane $( M_{\Frm},\.k )$ for $q = 0.5$ assuming an instantaneous transition ($\delta = 0$). Two regimes appear for the constraints, with the vertical ($k$-independent) bands at high PBH mass coming from PBHs in the SC phase and the $k$-dependent, curved regions originating from the evaporation during the MB phase. Notably, the constraints from the MB phase scale very closely with the PBH lifetime, which is displayed with the black contours for MB lifetimes $\tau = t_{\rm mb} \in [10^{-10},\.10^{40}]$\,s. In the upper white region, PBHs can make up the entire DM, whereas they are fully evaporated and unconstrained in the lower left region below the bound from BBN. 

As first demonstrated in Refs.~\cite{Montefalcone:2025akm, Dvali:2025ktz}, a non-instantaneous transition from the SC phase to the MB phase can significantly affect the constraints unless $\delta \lesssim 10^{-10}$~\cite{Montefalcone:2025akm}. This is illustrated in Fig.~\ref{fig:Mkplot-transitions}, which shows two illustrative cases: $q = 0.5$ with $\delta\sim\mathcal{O}(0.1)$ (left) and $q = 1$ with $\delta = 0$ (right). Note in particular that according to Eq.~(\ref{eq:transition}), a smooth transition with $\delta\sim\mathcal{O}( 0.1 )$ is more natural when considering $q\sim0.5$ than the instantaneous one (illustrated in Fig.~\ref{fig:lifetimes}). Comparing Fig.~\ref{fig:lifetimes} (assuming $q\sim0.5$ and $\delta = 0$) to the left panel of Fig.~\ref{fig:Mkplot-transitions} (assuming $q \sim 0.5$ and $\delta \sim \mathcal{O}( 0.1 )$), we see that the impact of a smoother transition is to exclude the entire region above $t_{\rm mb} \sim 1$\,s with observational constraints. This strongly limits low-mass PBHs with $M_{\Frm} < 10^{17}$\,g as candidates for all DM as discussed in Refs.~\cite{Montefalcone:2025akm, Dvali:2025ktz}. Furthermore, if the onset of MB occurs very early, with $( 1 - q )\ll 10^{-10}$, it implies $\delta \ll 10^{-10}$ according to Eq.~(\ref{eq:transition}). In the latter case, the constraints from the SC phase (vertical band) vanish, leaving only the excluded curved regions. This is illustrated in the right panel of Fig.~\ref{fig:Mkplot-transitions} for $q = 1$ and $\delta = 0$, which corresponds to a MB-only phase with instantaneous transition.

Importantly, the region of the parameter space that we focus on in this paper, where PBHs are fully evaporated and unconstrained, remains very similar. This corresponds to the region below the contour $t_{\rm mb} = 1$\,s. The BBN bound (blue area) only shifts by a factor $q = 0.5$ in PBH mass. Together with the earlier discussion in Section~\ref{sec:SC+MB}, this means that, for our analysis, the transition is not expected to strongly affect our conclusions. For this reason, we simply assume an instantaneous transition in the rest of this work.

%%%%%%%%%%%%%%%%%%%%%%%%%%%%%%%%%%%%%%%
\section{Dark Matter Abundance}
%\section{DM abundance}
\label{sec:DMabund}
%%%%%%%%%%%%%%%%%%%%%%%%%%%%%%%%%%%%%%%

The DM number density at evaporation results from the BH density at formation. To obtain a monochromatic BH distribution, we assume that the inflaton decays into radiation with an overdensity on a suitably small scale. Following the literature, we denote the initial PBH abundance as~\cite{Carr:2020gox}
\begin{equation}
    \beta
        \equiv 
            \Omega_{\rm PBH}(t_{ \Frm}) \,,
\end{equation}
which allows us to express the initial PBH number density as
\begin{eqnarray}
    n_{\mathrm{BH}}(t_\Frm)
        &=&
            \frac{\beta}{ M_{\Frm}}\rho_\Frm
        =
            3\.\beta\.(4 \pi\mspace{1mu}\gamma)^{2}\,T_\Frm^{3} \,,
\label{eq:nBHtF}
\end{eqnarray}
where $\rho_\Frm$ denotes the total energy density at the time of PBH formation. In particular, $\gamma$ captures the efficiency of collapsing the overdense region into the PBH with $M_{\Frm} = \gamma\.\rho_\Frm\.4\pi/3 H_\Frm^{-3}$ where $H_\Frm$ denotes the Hubble rate at formation~\cite{Carr:1975qj}. In this work, we use $\gamma = 1$ for concreteness. Notice that the value of $\gamma$ is usually reported to be smaller than one at BH formation, depending on the details of the gravitational collapse, i.e.~the mass of the PBH at formation is typically smaller than the horizon mass. However, in our case, we need to consider the final value including accretion, which typically induces $\gamma\gtrsim 1$ depending on the shape of the density fluctuations at the time of formation~\cite{Escriva:2021aeh}.

The full set of evolution equations taken into account in our numerical analysis contains the following Boltzmann equations:
\begin{eqnarray}
    \frac{{\rm d}\rho_{\rm BH}}{{\rm d}t}
        &=&
            -3 H \rho_{\rm BH}-\frac{\rho_{\rm BH}}{M}\frac{{\rm d}M}{{\rm d}t}\\[2mm]
    \frac{{\rm d}\rho_{\rm r}}{{\rm d}t}
        &=&
            -4 H \rho_{\rm r}+ \frac{\rho_{\rm BH}}{M}\left.\frac{{\rm d}M}{{\rm d}t}\right|_{\rm r}\\[2mm]
    \frac{{\rm d}\rho_{\rm DM}}{{\rm d}t}
        &=&
            -3 H \rho_{\rm DM}+\frac{\rho_{\rm BH}}{M}\left.\frac{{\rm d}M}{{\rm d}t}\right|_{\rm DM}\,,
    \label{eq:Boltz}
\end{eqnarray}
where ${\rm d} M / {\rm d}t|_J$ with $J\in \{\rm r,DM\}$ gives the BH mass evaporation rate into either radiation or DM as evaluated with {\tt BlackHawk} and $\rho_J$ with $J\in\{\rm r,DM,BH\}$ denotes the respective energy density. We also have to consider the Friedmann--Lema\^{i}tre equation for the Hubble expansion rate
\begin{equation}
    H^{2}(a)
        =
            \frac{8\pi}{3M_{\rm P}^{2}}\sum_J\rho_J(a)\,,
    \label{eq:Fried}
\end{equation}
where the sum runs over all species, including radiation, DM, a cosmological constant, as well as PBHs, while $\rho_J(a)$ captures the time (or scale-factor) dependence of the different energy densities in the Boltzmann Eqs.~(\ref{eq:Boltz}), and $H(a) = \text{d}\ln a/\text{d}t$. Note in particular that one can relate the energy density at formation to the Hubble rate at formation, $H_{\Frm} = (2t_\Frm)$, through the Friedmann--Lema\^{i}tre equation in a radiation dominated era, which implies that the formation time scales as the BH mass at formation and reads
\begin{equation}
    t_\Frm
        =
            \frac{M_{\Frm}}{\gamma\.M_{\rm P}^{2}}\,.
\end{equation}

Depending on the PBH fraction at formation, PBHs might come to dominate the Universe. The critical value of $\beta$ for which the PBHs come to dominate the Universe is estimated with $\beta_{\rm c} = a_{\Frm}/a_{\rm eq} = \sqrt{t_\Frm/t_{\rm eq}}$, where the subscript `${\rm eq}$' denotes the time of equality between the PBH-dominated and radiation-dominated (RD) phases and $a( t ) \propto t^{1/2}$ is assumed until equality. In particular, for PBH domination at the end of the SC phase, i.e.~for $t_{\rm eq} = t_q$, the critical value is given by
\begin{equation}
    \beta^{\rm sc}_{\rm c}
        =
            \sqrt{\frac{t_\Frm}{t_q}} =  \sqrt{\frac{3\.e_T}{\gamma(1-q^{3})}}\;\frac{M_{\rm P}}{M_{\Frm}}\,.
    \label{eq:betacsc}
\end{equation}
In contrast, in case of PBH domination at the end of the MB phase, i.e.~for $t_{\rm eq} = t_{\rm mb}$, we have 
\begin{equation}
    \beta^{\rm mb}_{\rm c}
        =
            \sqrt{\frac{t_\Frm}{q^{2}\.t_{\rm mb}}} = \sqrt{\frac{\kappa\.e_T}{\gamma(4\pi)^k}}\left(\frac{M_{\rm P}}{M_{\Frm}}\right)^{\!k+1}\frac{1}{q^{k+5/2}}\,.
    \label{eq:betacmb}
\end{equation} 
Note that this expression is valid if $t_{\rm mb}\gg t_q$. The extra factors of $1/q$ enter, as the abundance of PBHs from the MB phase is given by $q\.\beta$ as $(1-q)\.\beta $ has already evaporated away. In the case where PBH domination sets in during the SC phase, and for $t_{\rm mb}\gg t_q$, two distinct periods of PBH domination will happen (one in the SC and the other in the MB phase), separated by a brief RD phase. In the next subsections, we provide analytic estimates of the DM relic density assuming that the Universe is either radiation dominated at the time of DM production ($\beta<\beta_{\rm c}$) or matter dominated ($\beta>\beta_{\rm c}$), allowing to extract easily the time dependence of the expansion rate. 
 
The DM density will also depend on the scale factors at the different key moments. The scale factor at evaporation $a_\e$ depends on the assumptions about the MB phase. In particular, in the SC limit, assuming that $t_\text{ev} = t_{\rm sc}(M_{\Frm})$ and RD all along, we have a scale factor at evaporation 
\begin{equation}
    a_{\rm sc}
        =
            \; 3.3\times 10^{-31} \left(\frac{M_{\Frm}}{M_{\rm P}}\right)^{\!3/2}\,, 
\end{equation}
while, when the MB effect sets in, the SC phase has taken place up until
\begin{equation}
    a_q
        =
            a_{\rm sc }\big( 1-q^{3} \big)^{1/2}\,,
\end{equation}
where $a_q$ refers to the scale factor at time $t_q$, the factor $(1-q^{3})$ results from Eq.~(\ref{eq:tq}) and the power $1/2$ results from the RD assumption. On the other hand, when considering full evaporation in a MB phase (when we assume that $t_\text{ev} = t_{\rm mb}$), the scale factor at evaporation gets a factor of $k/2$ powers of entropy when assuming RD. Considering that $t_{\rm mb}\gg t_q$, we get 
\begin{eqnarray}
    a_{\rm mb}
        &=&
            a_{\rm sc}\,\sqrt{3}\,q^{3/2}\.S(q M_{\Frm})^{k/2}\,\\[2.5mm]
        &=&
            3.3\times 10^{-31} \sqrt{\frac{3}{\kappa}}\,(4\pi)^{k/2}\left(\frac{q\.M_{\Frm}}{M_{\rm P}}\right)^{\!k+3/2}\,,
    \label{eq:amb}
\end{eqnarray}
which denotes the scale factor at full evaporation at the end of the MB phase, as expected from Eq.~(\ref{eq:tautild}), and the second equality was obtained assuming that full evaporation occurs before the electroweak phase transition.  A full evaporation between the electroweak scale and BBN would give rise to an up to 25\% increase of the prefactor due to the change in the number of relativistic degrees of freedom in the early Universe.

We can now proceed to some analytic estimate for the relic abundance of light DM, $m_{\rm DM} < T_{\Frm}$, from PBH evaporation. Barring additional factors causing non-standard expansion, once BH evaporation has concluded, the energy density of DM at rest, $n_{\mathrm{DM}}\,m_{\mathrm{DM}}$, scales as $a^{-3}$ as expected from Eq.~(\ref{eq:Boltz}). We have thus
\begin{equation}
    \Omega_{\mathrm{DM}}(t_0)
        =
            \frac{m_{\mathrm{DM}}\.n_{ \rm DM}(t_\e)}{\rho_{\rm c}} \left( \frac{ a_\e }{ a_0 }\right)^{\!3}\,,
\label{eq:omdm0}
\end{equation}
with $n_{ \rm DM}(t_\e)$ being the DM number density, $t_\e$ denotes the time of full evaporation, $\rho_{c}$ is the critical energy density today, and $a_0 \equiv 1$ is the scale factor today. The DM number density at any time $t$ can be evaluated from the DM density at evaporation time as
\begin{eqnarray}
    n_\dm( t )
        &=&
            N_\dm(t_\e)\,n_\bh(t_\e)\mspace{-2mu}
            \left(\frac{a_\e}{a( t )}\right)^{\!3}\, ,
\end{eqnarray}
with $a_\e$ and $N_\dm(t_\e)$ corresponding, respectively, to the scale factor and the total number of DM particles produced through the BH evaporation at $t_\e$. The BH number density at full evaporation reads
\begin{equation}
    n_\bh(t_\e)
        =
            n_\bh(t_{\Frm})\mspace{-2mu}
            \left(\frac{a_{\Frm}}{a_\e}\right)^{\!3} = T_{\Frm}^{3}\,{\cal N}_\e\, ,
  \label{eq:nDM}
\end{equation}
where we introduced ${\cal N}_\e$, the time-independent BH number density variable at evaporation rescaled by the BH formation temperature. This quantity depends on the type of evaporation, and in particular on whether the latter happens in a matter-dominated (MD) or RD era. In the next subsection, we provide an analytical approximation of ${\cal N}_{\rm ev}$  (see Eq.~(\ref{eq:prefMDRD})). 

Note that the numerical results that we will provide are obtained using the full treatment of the Boltzmann equations instead of the analytical approximation.  

\begin{comment}
, without burst,
\begin{eqnarray}
a_{\rm mb}& = & a_{\rm sc}\,\sqrt{3}q^{3/2}S^{k/2}(q M_{\Frm})\,\\
& = & 3.3\times 10^{-31} \sqrt{3}(4\pi)^{k/2}\left(\frac{q\.M_{\Frm}}{M_{\rm P}}\right)^{k+3/2} \quad{\rm [no\, burst]}
\end{eqnarray}
while, with burst, we get
\begin{eqnarray}
a_{\rm mb}& = & a_{\rm sc}\,q^{3/2}\frac{S^{k/2}(q M_{\Frm})}{\sqrt{1+2k/3}}\,\\
& = & 3.3\times 10^{-31}  \sqrt{\frac{(4\pi)^k}{1+2k/3}}\left(\frac{q M_{\Frm}}{M_{\rm P}}\right)^{k+3/2} \quad{\rm [burst]}
\end{eqnarray}
\label{eq:aevsc}
as expected from Eq.~(\ref{eq:tautild}). 
\end{comment}

%%%%%%%%%%%%%%%%%%%%%%%%%%%%%%%%%%%%%%%
\subsection{The Semi-Classical Limit}
\label{sec:DMabundSC}
%%%%%%%%%%%%%%%%%%%%%%%%%%%%%%%%%%%%%%%

Estimating the ratios of scale factors between formation and SC evaporation, the DM relic abundance is given by~\cite{Baldes:2020nuv}
\begin{eqnarray}
    \Omega^{\rm sc}_{\mathrm{DM}}(t_0)
        &=&
            \frac{m_{\mathrm{DM}} N_{\mathrm{DM}} T_{\Frm}^{3}}{\rho_{\rm c}}\times {\cal N}_{\rm ev}\times a_{\rm sc}^{3} \,,
  \label{eq:OmDM1}
\end{eqnarray}
where ${\cal N}_{\rm ev}$, introduced in Eq.~(\ref{eq:nDM}), reads
\begin{equation}
	{\cal N}_{\rm ev}
        \simeq
            \begin{dcases*}
                3\,\beta\,(4\pi)^{2}\,\gamma^{1/2}\,(3\.e_T)^{3/2}\,\frac{M_{\rm P}^{3}}{M_{\Frm}^{3}}
                    & if $\beta<\beta_{\rm c}$\,,\\[1mm]
                3\,(4\pi)^{2}\,(4\.e_T)^{2}\,\frac{M_{\rm P}^{4}}{M_{\Frm}^{4}}
                    & if   $\beta>\beta_{\rm c}$\,.
            \end{dcases*}
  \label{eq:prefMDRD}
\end{equation}
Equivalently, we have
\begin{equation}
    \frac{ \Omega^{\rm sc}_{\mathrm{DM}}(t_0)\.h^{2} }{ 0.12 }
        =
            \left(\frac{m_{\mathrm{DM}} }{ 1\,\rm  GeV}\right) \times
            \begin{dcases*}
                \left(\frac{ M_{\Frm}}{7.1 \times 10^{12} M_{\rm P} }\right)^{1/2} \left( \frac{\beta}{2.2\times 10^{-14}}\right) 
                    & if  $\beta<\beta_{\rm c}$\,,\\[2mm]
                \left(\frac{ M_{\Frm}}{ 7.1 \times 10^{12} M_{\rm P} }\right)^{-1/2}
                    &  if   $\beta>\beta_{\rm c}$\,.
            \end{dcases*}
\label{eq:OmDM2}
\end{equation}

%%%%%%%%%%%%%%%%%%%%%%%%%%%%%%%%%%%%%%%
\subsection{The Case of a Long Memory-Burden Phase}
%\subsection{Long MB case}
\label{OmlongMB}
%%%%%%%%%%%%%%%%%%%%%%%%%%%%%%%%%%%%%%%

When the parameters are such that $t_{\rm mb}>t_0$ (corresponding to the upper right region in Figs.~\ref{fig:lifetimes} and~\ref{fig:Mkplot-transitions}), the DM abundance is
\begin{equation}
    \Omega_{\mathrm{DM}}(t_0)
        =
            \big(1-q^{2}\big)\.\Omega_{\mathrm{DM}}^{\rm sc}(t_0) +  \Omega_{\mathrm{BH}}^{\rm relic}(t_0)\,,
  \label{eq:OmDMlongMB}
\end{equation}
where the first term arises from the PBH SC evaporation phase, with the $(1-q^{2})$ weighting factor as in Eq.~(\ref{eq:Nq}), while the left over abundance of non-evaporated PBHs with mass $q M_{\Frm}$ is given by
\begin{equation}
    \rho_{\rm BH}^{\rm relic}
        =
            q M_{\Frm} \times n_{\rm BH}(T_{\Frm},M_{\Frm})\times (a_{\rm F}/a_0)^{3}\,,
\end{equation}
which yields
\begin{equation}
    \frac{\Omega_{\mathrm{BH}}^{\rm relic}(t_0)h^{2}}{0.12}
        =
            \frac{q}{0.5} 
            \left(\frac{ M_{\Frm}}{7.1\times 10^{12}\, M_{ P}}\right)^{\!-1/2} \frac{\beta}{1.4\times 10^{-21}}\quad \text{if}\,\beta<\beta_{\rm c}\,,
\label{eq:OmBHlongMB}
\end{equation}
i.e.~the remaining BHs for the range of masses of interest would largely overclose the Universe except for very low initial abundances $\beta$. Actually, the case of PBH domination ($\beta >\beta_{\rm c}$) is excluded as PBH would strongly overclose the Universe. Furthermore, assuming a smooth transition as discussed in Sec.~\ref{sec:SC+MB}, the region of $t_{\rm mb}>t_0$ would get strongly disfavoured by observational constraints, as illustrated in the left panel of Fig.~\ref{fig:Mkplot-transitions}, see Refs.~\cite{Montefalcone:2025akm, Dvali:2025ktz}.

%%%%%%%%%%%%%%%%%%%%%%%%%%%%%%%%%%%%%%%
\subsection{The Case of a Short Memory-Burden Phase}
\label{OmshortMB}
%%%%%%%%%%%%%%%%%%%%%%%%%%%%%%%%%%%%%%%

When the parameters are such that $t_{\rm mb} < t_{\rm BBN}$, all the DM is made out of the two evaporation phases. The relative contributions from the two phases are given by
\begin{align}
    \Omega_{\rm DM}^{\rm sc}(t_0)
        &=
            \frac{1-q^2}{1+(\xi-1) q^2}\Omega_{\rm DM}(t_0)\,\\
    \Omega_{\rm DM}^{\rm mb}(t_0)
        &=
            \frac{\xi q^2}{1 + (\xi-1)q^2}\Omega_{\rm DM}(t_0)\,,
    \label{eq:shortMBOmDM}
\end{align}
where the total DM relic abundance $\Omega_{\rm DM}(t_0)$ is given by (see also Ref.~\cite{Haque:2024eyh})\footnote{Note that our prefactors in Eq.~(\ref{eq:OmDMshortmb}) differ from those in Ref.~\cite{Haque:2024eyh} due to our treatment of greybody factors with {\tt BlackHawk}, instead of the geometric optics approximation.}
\begin{eqnarray}
    \frac{ \Omega_{\mathrm{DM}}(t_0)h^{2} }{ 0.12 }
        &\approx&
            \left(1+(\xi-1)q^2\right)\left(\frac{m_{\mathrm{DM}} }{ 1\,\rm GeV}\right) \nonumber\\[2mm]
        &&\times
            \begin{dcases*}
                1.7 \times \left(\frac{ M_{\Frm}}{M_{\rm P}}\right)^{1/2} \left( \frac{\beta}{10^{-7}}\right) 
                    & if  $\beta<\beta_{\rm c}$\,,\\
                1.3\times 10^{6} \times q^{-2} \left(\frac{\kappa}{(4\pi)^k}\right)^{1/2}\left(\frac{q M_{\Frm}}{M_{\rm P} }\right)^{\!-k-1/2}
                    &  if   $\beta>\beta_{\rm c}$\,.
            \end{dcases*}
\label{eq:OmDMshortmb}
\end{eqnarray}
As in the SC case, for DM production in a RD era ($\beta<\beta_{\rm c}$), the DM relic abundance scales as $M_{\Frm}^{1/2}$ and is independent of the MB parameters $q$ and $k$. This can be expected as the total number of produced particles does not depend on the particularities of the PBH evaporation dynamics as mentioned in, e.g., Ref.~\cite{Haque:2024eyh}. Note however that the value of $\beta_{\rm c}$ depends on the MB parameters as visible in Eq.~(\ref{eq:betacmb}). Furthermore, in the PBH-dominated era ($\beta>\beta_{\rm c}$), the DM relic abundance scales as $(q M_{\Frm}/M_{\rm P})^{-(2k+1)/2}$ displaying as strong dependence on $k$ as we consider $M_{\Frm}> M_{\rm P}$ and a more moderate dependence in $q$.

\section{Particle Dark Matter from Evaporation}
\label{sec:DMevap}
%%%%%%%%%%%%%%%%%%%%%%%%%%%%%%%%%%%%%%%

Sourcing the DM relic abundance from BH decays has been considered in a number of previous studies, see e.g.~\cite{Matsas:1998zm, Bell:1998jk, Khlopov:2004tn, Fujita:2014hha, Allahverdi:2017sks, Lennon:2017tqq, Baumann:2007yr, Morrison:2018xla, Hooper:2019gtx, Masina:2020xhk, Hooper:2020evu, Baldes:2020nuv}. Given an initial BH mass, $M_{\Frm}$, there generally exist two solutions to match onto the relic abundance, depending on whether the initial BH temperature, $T_{\Frm}$, is above or below $m_{\rm DM}$, see, e.g., Refs.~\cite{Fujita:2014hha, Lennon:2017tqq}. Here we focus on light DM particles ($m_{\mathrm{DM}}\ll T_{\Frm}$) emitted ultra-relativistically from PBH evaporation that might leave a NCDM~\cite{Lesgourgues:2011rh, Murgia:2017lwo} imprint on cosmological observables. In the case of $m_{\rm DM} \gg T_{\Frm}$, the DM ends up cold enough to not be subject to structure-formation constraints~\cite{Fujita:2014hha} and will be of no further interest to us here.

%%%%%%%%%%%%%%%%%%%%%%%%%%%%%%%%%%%%%%%
\subsection{Non-Cold Dark Matter Phase-Space Distribution}
\label{sec:NCDM-PSD}
%%%%%%%%%%%%%%%%%%%%%%%%%%%%%%%%%%%%%%%

In order to estimate the effect of NCDM from evaporating PBH, we shall study in more detail the form of the momentum distribution. Here we will consider that the energy of the emitted particles is momentum dominated ($E\simeq p$). Let us introduce a NCDM temperature
\begin{equation}
    \Ts( t )
        =
            T_\star(t_\e)\.\frac{a(t_\e)}{a( t )} \,,  
\label{eq:Tncdm}
\end{equation}
with $T_\star(t_\e) = T_{\Frm}$ for the SC phase while $T_\star(t_\e) = T_{\Frm}/q$ for the MB phase. We also introduce a time-independent dimensionless momentum variable
\begin{equation}
    {\tt q}
        =
            \frac{p( t )}{\Ts( t )}\,,
  \label{eq:qncdm}
\end{equation}
where $p( t )$ is the proper momentum and ${\tt q}$ denotes the rescaled comoving momentum for NCDM particles that we choose to use in {\sc class}.

We want to evaluate the DM momentum distribution, $f_\dm({\tt q})$. The latter is related to the DM number density at $t_\e$ as $n_\dm (t_\e) = T_\star(t_\e)^{3}\int {\rm d}^{3}{\tt q}\;g_\dm\,f_\dm ({\tt q}) / (2\pi)^{3}$ and takes the form
\begin{eqnarray}
    \frac{g_\dm}{(2\pi)^{3}}\,{\tt q}^{2} f_\dm({\tt q})
        & = &
            {\cal N}_\e\,T_\star(t_\e)\int_{t_{\Frm}}^{t_\e}{\rm d}t\,\frac{a_\e}{a}\,\frac{\mathrm{d}N_\dm }{\mathrm{d}p\.\mathrm{d}t}\,, \label{eq:fDM}
\end{eqnarray}
after full evaporation. Note that the number of evaporated particles per unit time and momentum has to be evaluated in the right phase and that a ratio of scale factors in the time integral has to be introduced to account for redshifting of momenta during a non-instantaneous evaporation~\cite{Lennon:2017tqq}. For further discussion, we introduce the rescaled DM momentum distribution
\begin{equation}
      \tilde f_\dm({\tt q})
        =
            \frac{1}{ {\cal N}_\e}\frac{T_\star(t_\e)^{2}}{M_{\rm P}^{2}}  \,\frac{1}{(2\pi)^{3}}\,{\tt q}^{2} f_\dm({\tt q})\,,
      \label{eq:tildf}
\end{equation}
where the prefactor ${T_\star(t_{\rm ev})^{2}}/(M_{\rm P}^{2} {\cal N}_\e)$ has been introduced in order to extract a universal momentum distribution, $\tilde f_\dm({\tt q})$, that is relatively\footnote{This is only valid for RD, for $T_{\Frm}>T_{\rm EW}$.} BH mass independent, see, e.g., Refs.~\cite{Lennon:2017tqq, Decant:2021mhj}. Furthermore, the moments of the distribution are defined as
\begin{equation}
    \big\langle {\tt q}_\dm^n \big\rangle
        =
            \frac{\int {\rm d}{\tt q}\,{\tt q}^n\times T_\star(t_\e)\,\tilde f_\dm({\tt q})}{\int {\rm d}{\tt q} \,\tilde f_\dm({\tt q})}\,.
  \label{eq:meanp}
\end{equation}
In particular, considering $q = 0.5$  and using {\tt BlackHawk}, after full evaporation, we get
\begin{alignat}{5}
    \langle {\tt q}_\dm \rangle^{}_{\rm sc}
        & =  4.2&
            \quad    \sqrt{\langle {\tt q}^{2}_\dm \rangle^{}_{\rm sc}}
        & =  5.1&   
        &\text{SC phase}\, ,\\[2mm]
    \langle {\tt q}_\dm \rangle^{}_{\rm mb}
        & =  2.8\,(7.6)&
            \sqrt{\langle {\tt q}^{2}_\dm\rangle^{}_{\rm mb}}
        & =  3.2 (\infty)\quad&
        & \text{MB phase without (with) burst}\, .
\end{alignat}
Note that the $\langle {\tt q}^{2}_\dm \rangle_{\rm sc}^{1/2}$ would be infinite for $q = 0$ (pure SC) as the DM distribution function scales as ${\tt q}^{-3}$ for large ${\tt q}$ values, meaning that it diverges logarithmically when computing the average, see also the discussion in Ref.~\cite{Ballesteros:2020adh}. Also note that $k$ affects the position of the peaks in the distribution but not its overall shape, so that the moments of the distribution are not affected by its value. 

\begin{figure}[t]
    \centering
    \includegraphics[width = 0.68\linewidth]{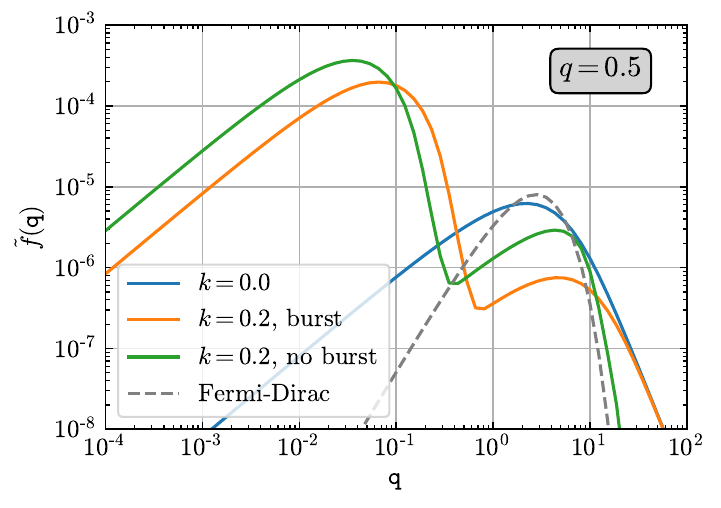}
    \caption{Rescaled momentum distributions  as a function of the comoving momentum ${\tt q}$. Continuous lines represent the NCDM distribution $\tilde{f}(q)$ for $k = 0$ in blue, in which case we recover the SC behaviour, and $k = 0.2$ in the case of burst in orange and of no burst in green. In the latter cases, we have set  $q = 0.5$ or equivalently we consider that MB sets in at half mass. A rescaled Fermi-Dirac distribution is also shown for reference with a dashed gray line.}
    \label{fig:tidf2phases}
\end{figure}

In Fig.~\ref{fig:tidf2phases}, we show the rescaled DM momentum distribution considering one or two evaporation phases.  In the figure,  the dashed gray line is shown for reference. It is a Fermi-Dirac (FD) distribution with a normalization and temperature fixed to have a maximum around the same rescaled momentum than for PBH evaporation in a SC phase, represented with a blue line ($k = 0$ case). The orange and green curves show the SC+MB case when considering $q = 0.5$ and $k = 0.2$. The latter value is chosen in such a way that the two peaks in the distribution can appear on the plot for illustrative purposes. Let us emphasize though that such behaviour only appears in the bottom right corner of Fig.~\ref{fig:lifetimes} when $a_{\rm mb}/a_q\sim 1$ where $k<0.5$, a region of the parameter space that would be disfavoured by theory argument in Ref.~\cite{Dvali:2020wft}. Note that Fig.~\ref{fig:tidf2phases} has been generated assuming a RD era during DM production through PBH evaporation. For illustration, we show in the Appendix, in Fig.~\ref{fig:tidf2phasesMD}, how the rescaled distributions would change if DM production would happen in a MD era during PBH evaporation.

On general grounds, it is interesting to note that the BH could go through two phases of evaporation, giving rise to a multimodal DM momentum distribution after full evaporation The MB without burst induces a sharper cutoff of the distribution at large momenta. With burst, it is a power-law tail (as in the SC case) while without burst it is an exponential cut-off. Also, the burst versus no burst peaks are shifted with respect to each other, because when evaporation takes place with a burst it happens faster, see Eq.~(\ref{eq:tautild}). 

%%%%%%%%%%%%%%%%%%%%%%%%%%%%%%%%%%%%%%%
\subsection{Non-Cold Dark Matter Constraints}
\label{sec:WDM}

Light DM particles ($m_{\mathrm{ DM}}\ll T_{\Frm}$) emitted ultra-relativistically from PBH evaporation will behave as NCDM~\cite{Lesgourgues:2011rh, Murgia:2017lwo} when it comes to their imprint on cosmological observables. Such DM candidates are expected to leave an imprint on Lyman-$\alpha$ forest observations due to their free-streaming properties, suppressing small-scale structures. Indeed, after re-ionization, residual neutral hydrogen gas in the intergalactic medium (IGM) produces a series of absorption lines in the spectrum of distant quasars referred to as the Lyman-$\alpha$ forest, through scattering in the Lyman-$\alpha$ transition. The fraction of light that is absorbed, the transmitted flux, depends on the small-scale structure properties along the line of sight. The flux power spectrum of such distant quasars can thus give information about the matter distribution and potentially the free-streaming properties of DM on the probed scales, see e.g. Refs.~\cite{Viel:2005qj, Viel:2013fqw}. 
 
Multiple approaches in the literature have been used to translate the derived constraints computed for thermal WDM{\,---\,}the NCDM archetype that is produced through the freeze-out mechanism while relativistic{\,---\,}into constraints for any NCDM candidate.  In this paper in particular, we will follow~\cite{Murgia:2017lwo}, who proposed a methodology, using the so-called area criterion (see Sec.~\ref{sec:areacrit}), to translate the 95\%CL on the WDM mass of
\begin{equation}
    m_{\rm WDM}^{\text{Ly}-\alpha}
        =
            5.3\,{\rm keV}
    \label{eq:mlya}
\end{equation}
derived in Ref.~\cite{Irsic:2017ixq} with the datasets of XQ-100 and HIRES/MIKE. Note that the more recent analysis of Ref.~\cite{Irsic:2023equ} has slightly improved this bound using HIRES and UVES data. We also provide a less conservative estimate of the Lyman-$\alpha$ bound from a velocity-dispersion estimate, as introduced in Ref.~\cite{Ballesteros:2020adh}, see Sec.~\ref{sec:warmcrit} and further discussion in Sec.~\ref{sec:areacrit}.

A particularity of our analysis for NCDM from PBH evaporation is that we expect to always encounter at least two populations of DM arising from the two stages of PBH evaporation. This is in contrast to previous analyses that estimate bounds on NCDM from PBH evaporation (see, e.g., Refs.~\cite{Fujita:2014hha, Lennon:2017tqq, Morrison:2018xla, Baldes:2014rda, Masina:2020xhk, Cheek:2021odj, Cheek:2022mmy, Barman:2024iht}) as these considered that all of the DM is made up of a single NCDM population. Below, we first provide more details on the properties of these two DM populations and then describe how we proceed to provide a conservative estimate of the Lyman-$\alpha$ constraint on such a scenario.

%%%%%%%%%%%%%%%%%%%%%%%%%%%%%%%%%%%%%%%
\subsubsection{Two Dark Matter Populations from Primordial Black Hole Evaporation}
%\subsubsection{Two Dark Matter Populations from PBH Evaporation}
\label{sec:twopop}
%%%%%%%%%%%%%%%%%%%%%%%%%%%%%%%%%%%%%%%

Considering PBH evaporation under the influence of MB effects, we will encounter two types of situations with (partial) contribution of NCDM from PBH evaporation:

\begin{enumerate}[wide, labelwidth = !, labelindent = 0pt]

    \item\label{lab:CWDM} \underline{\bf CDM + NCDM:} One could end up with a cold DM (CDM) and a NCDM population. This could happen in the long MB case, i.e. if the PBHs have gone through their first SC stage, giving rise to one population of NCDM particles, while the second MB stage has not yet concluded its evaporation. In the latter case, stable remnant PBHs, with long lifetime due to the MB effect, effectively contribute as CDM. Assuming an instantaneous transition, such a scenario would only happen for large masses and $k$ values ($k> 1$ and $10^{3} {\rm g} < M_{\Frm} < 10^{10}$\,g), corresponding to the top right corner of Fig.~\ref{fig:lifetimes}. Even in the latter case, the population of NCDM from the first SC stage would typically represent a suppressed contribution to the total DM abundance if $q> 0.5$, see the discussion in Sec.~\ref{sec:longMB}. Furthermore, considering a smoother transition, this parameter space region in $k$ and $M_{\Frm}$ is strongly constrained (see the discussion in Sec.~\ref{sec:obs-cons-PBH}). 
 
    Furthermore, assuming that PBHs have fully evaporated, a fraction of the particle DM population could behave as NCDM in case of a relatively early SC stage (giving rise to particle CDM) followed by a more recent MB stage (giving rise to NCDM). The momenta of the population of particle DM from the early SC stage would have had time to be redshifted sufficiently to behave as CDM. This scenario is expected to occur for lower PBH masses than in the case of long MB discussed above. From Eq.~(\ref{eq:Nq}), we see that for $q > 0.7$ the NCDM from the second stage of evaporation gives rise to more than half of the DM population, independently of the other PBH parameters. 
  
    \item\label{lab:2NCDM} \underline{\bf Two populations of NCDM:} When the MB and SC evaporation phases occur close to each other, two populations of particle NCDM from PBH evaporation, with different velocity dispersion, are expected to be present today. This is expected to happen for small $k$ values ($k<0.5$), see the region below the gray contour indicating $a_{\rm mb}/a_{\rm q} = 10$ in Fig.~\ref{fig:lifetimes}. In the latter case, one has to deal with multimodal NCDM velocity distributions as illustrated in Fig.~\ref{fig:tidf2phases} by the green (orange) curve for $k = 0.2$, assuming no burst (a burst). This is however expected to happen in a very restricted part of the parameter space that is actually disfavored by the theoretical arguments~\cite{Dvali:2020wft}.

\end{enumerate}
Note that for certain combinations of PBH ($M,k, q$ and $\beta$) and particle DM ($m_{\rm DM}$) parameters, we might not saturate the Planck relic DM abundance with the DM from PBH evaporation and the potential PBH remnants. In the latter cases, for concreteness, we assume that the rest of the DM is made of another CDM component.

The above NCDM scenarios are expected to give rise to a suppression of the matter power spectrum at small scales. The latter is well visible when displaying the ratio of the NCDM matter power spectrum with the CDM power spectrum, commonly referred to as the transfer function $T({\tt k})$, defined as
\begin{equation}
    T^{2}({\tt k})
        =
            P_{\rm NCDM}({\tt k})/P_{\rm CDM}({\tt k})\,,
    \label{eq:transferfn}
\end{equation}
where ${\tt k}$ denotes the wave number, Fourier dual of a comoving distance not to be confused with the MB parameter $k$, and $P_i({\tt k})$ denotes the matter power spectrum in a $i = $ CDM or NCDM cosmology. The suppression of the transfer function could be followed by a plateau in the cases where the NCDM produced through the above scenarios do not make 100\% of the DM content. 

\begin{figure}
    \centering
    \includegraphics[width=0.6\linewidth]{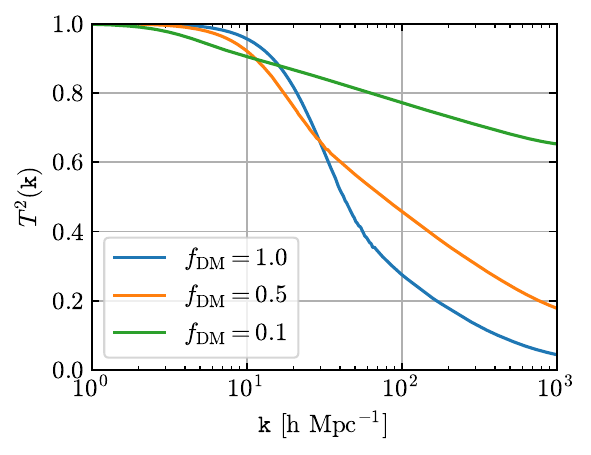}
    \caption{Transfer function $T^2(\texttt{k})$ for $k=2$, $q=0.5$, $\beta=10^{-16}$, $M_{\rm F}=2\times10^4\,M_{\rm P}$, and $m_{\rm DM}$ chosen such that DM from PBH evaporation accounts for a fraction of $f_{\rm DM}=1$, 0.5, and 0.10 of the entire DM. In this scenario, only the MB phase of evaporation contributes to the NCDM component and the parameters were chosen to be close to the bound from the Lyman-$\alpha$ forest, shown in Figure~\ref{fig:mdmvsMF}.}
    \label{fig:transferfn}
\end{figure}

In order to obtain the transfer functions, we have introduced the DM momentum distribution derived in Sec.~\ref{sec:NCDM-PSD} in the {\tt CLASS} code. Some example of the resulting transfer functions are shown in Fig.~\ref{fig:transferfn} for $q = 0.5$ and $k = 2$. The three color curves illustrate the cases of $\beta = 10^{-16}$ from PBH evaporation without burst with particle DM masses set to 3.3, 1.6 and $0.3 \times 10^6\,$GeV giving rise respectively to a fraction $f_{\rm DM}=1$ (blue curve), 0.5 (orange curve) and 0.1 (green curve) of the  DM relic abundance (the rest is assumed to behave as CDM).  Out of those particle DM contributions,  the NCDM arising from the MB phase would represent $2q^2/(1+q^2)\sim 40$, 20 and 4\% of the DM (see Eq.~(\ref{eq:shortMBOmDM})).  As in the case of Warm + CDM we see a plateau appearing at large $k$ values~\cite{Viel:2005qj, Boyarsky:2008xj, Lesgourgues:2011rh,Hooper:2022byl, Euclid:2024pwi}, especially when the fraction of NCDM is small (green-coloured curve).

Ideally, in order to account for Lyman-$\alpha$ constraints, we should make use of the full likelihood pipeline proposed in Ref.~\cite{Hooper:2022byl} with the velocity distributions derived in Sec.~\ref{sec:NCDM-PSD} as an input. However, such an analysis is computationally costly as the power spectrum computation can take a few minutes for a single set of PBH and DM mass parameters lying at the expected limit of the NCDM parameter space when considering the most accurate computation in {\tt CLASS} (involving {\tt ncdm\_fluid\_approximation = 3}). 

In this paper, we instead compare two different approaches. The first and simplest one compares the typical velocity dispersion of the NCDM candidates to the one of a constrained WDM particle, see Sec.~\ref{sec:warmcrit}. Similar approaches have been used in the literature (see, e.g.~Refs.~\cite{Decant:2021mhj, Baldes:2020nuv}). The second approach will involve the computation of the evaporated DM matter power spectrum making use of the velocity distributions derived above as input for the Boltzmann code {\tt CLASS}. Comparison with the case of thermal WDM will be made by using the area criterion introduced in Ref.~\cite{Murgia:2017lwo}, see Sec.~\ref{sec:areacrit}. Such an approach was argued to provide a good estimate of the Lyman-$\alpha$ forest constraints in the context of NCDM transfer functions that are strongly suppressed at large $k$ values, which is not always the case here (as visible in Fig.~\ref{fig:transferfn}). Note that the authors of Ref.~\cite{Murgia:2017lwo} emphasize that the approximation of the area criterion shall be used as a first quantitative step towards a more comprehensive analysis. The area criterion was further tested against several forms of distribution functions displaying a plateau at small scales in Ref.~\cite{Hooper:2022byl} in the transfer function. In the latter case, the authors mention that such a criterion is not a good estimate of the small-scale suppression. In Appendix~\ref{sec:WCDM}, we argue that such an approach is expected to provide conservative constraints on NCDM+CDM by comparing the results of Ref.~\cite{Hooper:2022byl} for WDM+NCDM with the area criterion estimate. For that reason, we will make use of the area criterion as a conservative estimate of the small-scale structure constraints in Sec.~\ref{sec:param-space}. Also note that for the choice of parameters made for the curves in Fig.~\ref{fig:transferfn}, the considered NCDM population saturates these constraints.

%%%%%%%%%%%%%%%%%%%%%%%%%%%%%%%%%%%%%%%
\subsubsection{Recasting from Velocity Dispersion}
\label{sec:warmcrit}
%%%%%%%%%%%%%%%%%%%%%%%%%%%%%%%%%%%%%%%

When NCDM arises from one single production mechanism or production channel and its velocity distribution is unimodal, it has been argued that its velocity dispersion today can be used to provide an estimate of the free-streaming constraints, see, e.g., Refs.~\cite{Lennon:2017tqq, Bae:2017dpt, Ballesteros:2020adh}. The idea is to compare the NCDM velocity dispersion today $\sigma_{\rm DM,0} = {\sqrt{\langle p^{2}\rangle_0}}/{m_{\rm DM}}$ with the one excluded for thermal WDM. In particular, for one NCDM candidate accounting for all the DM of the Universe, the DM velocity dispersion is given by
\begin{equation}
     \sigma_{\rm DM,0}
        =
            \sqrt{\langle {\tt q}_\star^{2}\rangle}\;
            {\tt T_{\rm NCDM}}\.
            \frac{T_\gamma(t_0)}{m_{\rm DM}} \qquad {\rm with}\qquad\,
    {\tt T_{\rm NCDM}}
        =
            a( t )\.\frac{T_\star( t )}{T_\gamma(t_0)}\,.
\label{eq:warmness}
\end{equation}
The dimensionless and time-independent parameter ${\tt T_{\rm NCDM}}$ compares the temperature scale relevant for the NCDM production, $T_\star ( t )$ defined in Eq.~(\ref{eq:Tncdm}), which scales as $a( t )^{-1}$, with the CMB temperature today, i.e.~at time $t_0$. ${\tt T_{\rm NCDM}}$ is used as defined in Eq.~(\ref{eq:warmness}) in {\tt CLASS}. In the case of thermal WDM, we have $\sqrt{\langle {\tt q}_\star^{2}\rangle} = 3.59$ for a Fermi-Dirac distribution when setting ${\tt T_{\rm NCDM}} = 0.16 ({\rm keV}/m_{\rm WDM})^{1/3}$.\footnote{For WDM, we set ${\tt T_{\rm NCDM}} = a_{\rm prod}\. T_\gamma(t_{\rm prod})/T_\gamma(t_0) = [ g_{\star\,s} (t_0)/g_{\star\,s} (t_{\rm prod}) ]^{1/3}$ where $\,g_{\star\,s} $ denotes the number of relativistic degrees of freedom contributing to entropy. In particular $\,g_{\star\,s} (t_{\rm prod}) = 956\,m_{\rm DM} /$keV is the number of relativistic degrees of freedom contributing to entropy at the time of WDM production, corresponding to WDM freeze-out, scales as $m_{\rm WDM}$ when accounting for all the DM, such that ${\tt T_{\rm NCDM}} = 0.16\,({\rm keV}/m_{\rm WDM})^{1/3}$.} Imposing $\sigma_{\rm DM,0}\lesssim \sigma_{\rm WDM,\.0}^{\text{Ly-}\alpha}$, where $\sigma_{\rm WDM,\.0}^{\text{Ly-}\alpha}$ is the velocity dispersion of a WDM candidate saturating some Lyman-$\alpha$ bound, the lower bound on some other NCDM candidate reads
\begin{equation}
    m_{\rm DM}
        \gtrsim
            1.74
                \,{\rm keV} \times \sqrt{\langle {\tt q}_\star^{2}\rangle}\;{\tt T_{\rm NCDM}}\times\left(\frac{m_{\rm WDM}^{\text{Ly-}\alpha}}{\rm keV}\right)^{\!4/3}\. ,
  \label{eq:Kamada}
\end{equation}
where $m_{\rm WDM}^{\text{Ly-}\alpha}$ is the thermal WDM candidate mass saturating the Lyman-$\alpha$ bound, see also Ref.~\cite{Ballesteros:2020adh, Decant:2021mhj, DEramo:2020gpr} for applications in other frameworks.
 
When multiple production channels are at the origin of the DM relic abundance, a first naive estimate of the Lyman-$\alpha$ bound in the case of mixed scenarios could be extracted by comparing again the total velocity dispersion, arising from all mechanisms of production, to the one of thermal WDM saturating the Lyman-$\alpha$ bound when $\Omega_{\chi} h^{2} = 0.12$. Within this framework, one gets
\begin{equation}
    m_{\rm DM}
        \gtrsim
            1.74 \,{\rm keV} \times \left(\frac{m_{\rm WDM}^{\text{Ly-}\alpha}}{\rm keV}\right)^{\!4/3}\!
            \times \left[\,\sum_{\rm prod} \left(\frac{\Omega_{\chi}\.h^{2}|_{\rm prod}}{0.12}\right) \times \Big(\big\langle {\tt  q_\star}^{2}\big\rangle\,{\tt T_{\rm NCDM}^{2}}\Big)\bigg|_{\rm prod}\right]^{1/2} \. ,
  \label{eq:Kamada-mix}
\end{equation}
where it has been assumed that $\Omega_{\chi} h^{2} = 0.12$ in order to compare to the thermal WDM constraints, where the sum runs over the production channels. In particular, in the case of DM production from SC+MB stages, we have
\begin{eqnarray}
\label{eq:Kamada-mixBH}
    m_{\rm DM}
        &\gtrsim&
            1.2 \,{\rm keV} \times \left(\frac{m_{\rm WDM}^{\text{Ly-}\alpha}}{\rm keV}\right)^{\!4/3} \sqrt{\frac{M_{\Frm}}{M_{\rm P}}}\frac{1}{\sqrt{1+(\xi-1)\.q^{2}}} \\ \nonumber
        &&\!\!
            \times\left[\big(1-q^{2}\big)\big(1-q^{3}\big) \big\langle {\tt q}^{2}_\dm \big\rangle_{\rm sc}+\left(\xi\.(1-q^{3})+\frac{3\xi}{\kappa}\,(4\pi)^k\,q^{3+2k}\left(\frac{M_{\Frm}}{M_{\rm P}}\right)^{\!2k} \right)\big\langle {\tt q}^{2}_\dm \big\rangle_{\rm mb} \right]^{1/2}\,.
\end{eqnarray}
Assuming that $t_{\rm mb}\gg t_{\rm sc}$ effectively implies that the first two terms are negligible. In particular, in the latter case, for $M_{\Frm} = 10 $\,g with $q = 0.5$ and $k = 0.5$, using Eqs.~(\ref{eq:meanp}) and~(\ref{eq:mlya}), we get a lower bound of $m_{\rm DM}\gtrsim 17$ GeV without burst. The latter is one order of magnitude larger than the estimate of 1.6 GeV reported in Ref.~\cite{Barman:2024iht} using a more approximate evaluation of the velocity dispersion.\footnote{Note in particular that~\cite{Barman:2024iht} considers evaporation with burst, in which case we would obtain an infinite result due the infinite second momenta in the latter case in the MB phase, see Eq.~(\ref{eq:meanp}). } The bound of Eq.~(\ref{eq:Kamada-mixBH}) essentially scales with the PBH mass and the MB parameter as $k$ as $ (M_{\Frm}/M_{\rm P})^{(2k+1)/2}$. We therefore expect an increasing lower bound on the DM mass for larger PBH mass and larger $k$ values. Both effects are quite natural as larger BH mass and larger $k$ values lead to full evaporation at a later time, leaving less time for the DM to get its momentum redshifted. Also, as already emphasized in e.g.~\cite{Baldes:2020nuv}, applying the Lyman-$\alpha$ bound to DM from evaporating BHs can exclude DM candidates with masses much higher than keV scale.

%%%%%%%%%%%%%%%%%%%%
\subsubsection{Recasting from Area Criterion}
\label{sec:areacrit}

An alternative approach to extract the Lyman-$\alpha$ bounds on NCDM scenarios is based on the area criterion introduced in Ref.~\cite{Murgia:2017lwo}, see also e.g.~Refs.~\cite{Schneider:2016uqi, DEramo:2020gpr,Egana-Ugrinovic:2021gnu, Decant:2021mhj} for applications. In order to quantify the suppression of the power spectrum in the NCDM model $X$, one computes the area estimator
\begin{equation}
    \delta A_X
        =
            \frac{A_{\rm CDM}-A_X}{A_{\rm CDM}} \quad {\rm with} \quad A_X = \int_{\rm {\tt k}_{\rm min}}^{{\tt k}_{\rm max}} \d {\tt k}'\,\frac{P_{1\mathrm{D}}^{X}({\tt k}')}{P_{1\mathrm{D}}^{\rm CDM}({\tt k}')}\,,
\end{equation}
where the 1D power spectrum in the DM scenario $X$ is obtained from the 3D power spectrum $P_X({\tt k})$ as $P^X_{1\mathrm{D}}({\tt k}) = \int_{\tt k}^{\infty} {\rm d}{\tt k}'\,{\tt k}'\,P_X({\tt k}')$. Reference~\cite{Murgia:2017lwo} associates a range of probed comoving scales between 
\begin{equation}
    {\tt k}_{\rm min}
        =
            0.5\.h/{\rm Mpc\qquad and \qquad} {\tt k}_{\rm max} = 20\.h/{\rm Mpc}
    \label{eq:krange}
\end{equation}
to the MIKE/HIRES+XQ-100 combined data set (see also Ref.~\cite{Irsic:2017ixq}). For the cosmological and precision parameters considered in our analysis, we get
\begin{equation}
    \delta A_{\rm WDM}
        =
            0.32\qquad {\rm for} \qquad m_{\rm WDM}
        =
            5.3\,{\rm keV}\,,
  \label{eq:dAw}
\end{equation}
where we report the threshold area criterion that would saturate the Lyman-$\alpha$ bound on WDM of 5.3 keV considered in Eq.~(\ref{eq:mlya}). A NCDM scenario, with one or more NCDM populations, that would give rise to $\delta A_X = \delta A_{\rm WDM}$ above is thus expected to saturate the WDM Lyman-$\alpha$ bound considered here. Larger $\delta A_X$, corresponding to stronger suppressions of the small-scale power spectrum, are excluded while smaller $\delta A_X$ are allowed.

\begin{figure}[t]
    \centering
    \includegraphics[width = 0.65\linewidth]{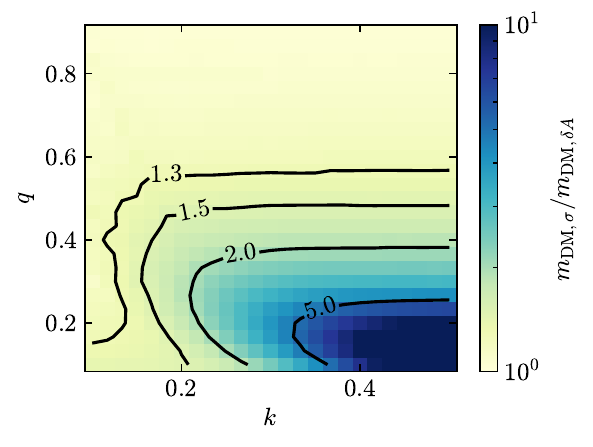}
    \caption{Ratio of the lower mass bound on NCDM  produced from PBH evaporation from velocity dispersion ($m_{\rm DM,\sigma}$) and from the area criterion ($m_{\mathrm{DM},\delta A}$). The ratio appears as the color gradient projected in the $(k,q)$ memory burden parameter space. Darker color indicates a larger discrepancy between the two criteria to evaluate the NCDM bound. This plot assumes that all the dark matter is produced from evaporating PBHs with $M_{\rm F}=10^2\,\rm{g}$. Note that the values for which the two criteria diverge significantly are not favoured by theory \cite{Dvali:2018xpy, Dvali:2021byy}.
    %Note for low $k$ and $q$ values the two phases MB and SC give NCDM with similar $T_{\rm ncdm}$ and the SC phase become more relevant giving rise to an ok estimate as $f_{\rm NCDM}^{\rm tot}\sim 1$. 
    }
    \label{fig:velovsarea}
\end{figure}

The area criterion is expected to provide a conservative estimate of the Lyman-$\alpha$ bound with respect to the velocity-dispersion estimate reported in Eq.~(\ref{eq:Kamada-mix}) in the case of W+CDM scenarios. This is particularly the case for small fractions of WDM,  see the discussion in App.~\ref{sec:WCDM}. The area criterion also provides a conservative bound when considering two populations of NCDM candidates arising from two phases of PBH evaporation instead of W+CDM. This is illustrated in Fig.~\ref{fig:velovsarea} where we project, with a gradient of colour, the ratio between the excluded NCDM mass from the velocity dispersion criterion from Eq.~(\ref{eq:Kamada-mix}), denoted as $m_{\rm DM,\.\sigma}$, and the one excluded by the area criterion saturating $\delta A$ reported in Eq.~(\ref{eq:dAw}), denoted as $m_{\mathrm{DM},\delta A}$, in the plane of $(q,k)$ for a fixed initial PBH mass of $M_{\Frm} = 10^{2}$\,g. We see that the ratio is always larger or equal to one, i.e.~$m_{\mathrm{DM},\delta A}<m_{\rm DM,\.\sigma}$, and becomes especially large in the region of lower $q$ values and larger $k$ values (bottom right part of the plot). %
%In the latter case, DM from the SC phase behaves as CDM and only a small fraction of NCDM from the MB phase is expected. Such scenarios would be excluded by the velocity-dispersion criterion but not by the area criterion. 
%In particular, we  see that the velocity-dispersion bound estimate becomes much stronger that the area criterion one when considering  $q < 0.5$, i.e.~when the PBH enters the MB phase while it has not yet lost half of its mass. 
Yet, such small $q$ values ($q < 0.5$) would not be favoured by theory arguments (see the discussion in Sec.~\ref{sec:MB}). Also note that the bottom left part of the plot focuses on the low-$k$ region ($k < 1$), illustrating the case of two NCDM populations (corresponding to the scenario \ref{lab:2NCDM} above) as SC and MB evaporation are expected to occur close in time, see also the gray contour in Fig.~\ref{fig:lifetimes}.

As the area criterion gives rise to the most conservative NCDM bound, we use the latter in the rest of our analysis to constrain DM from PBH evaporation.

%%%%%%%%%%%%%%%%%%%%%%%%%%%%%%%%%%%%%%%
\section{Parameter Space}
\label{sec:param-space}
%%%%%%%%%%%%%%%%%%%%%%%%%%%%%%%%%%%%%%%

In this paper, we are particularly interested in the NCDM imprint induced by DM produced through PBH evaporation with PBH masses below the BBN threshold, around $M_{\Frm} \sim 10^{10}$\,g in the SC approach. This will happen in two very different parts of the $(M_{\Frm},k)$ plane. As visible in Fig.~\ref{fig:lifetimes}, relatively large $k$ and $M_{\Frm}$ values give rise to a stable population of PBHs in their extended MB phase that will essentially saturate the DM abundance and behave as CDM. Yet, a small fraction of NCDM from PBH evaporation in the SC phase is obtained for small values of $q$. We briefly comment on the latter in Sec.~\ref{sec:longMB}.

On the other hand, for shorter MB phases, we should not spoil BBN constraints, i.e.~we have to impose $t_{\rm mb}< t_{\rm BBN}$ with $t_{\rm BBN}\sim 1$\,s refers to the time marking the onset of BBN. Furthermore, a lower bound on the evaporating PBH mass arises from Planck constraints on inflation~\cite{Akrami:2018odb} that translate into a lower bound on the PBH mass of $M_{\Frm} \sim 0.1$\,g. This constraint is independent of $\beta, k$ and $q$. Looking at Fig.~\ref{fig:lifetimes}, we see that this implies $k\lesssim 4$ for viable scenarios of short MB phases. In the latter case, we encounter two particle DM populations from PBH evaporation that could both behave as CDM, as NCDM, or behave as CDM+NCDM. The NCDM arising from the MB phase is always the warmer population. We detail more the latter case in Sec.~\ref{sec:shortMB}.

%%%%%%%%%%%%%%%%%%%%%%%%%%%%%%%%%%%%%%%
\subsection{Long Memory-Burden Phase}
\label{sec:longMB}
%%%%%%%%%%%%%%%%%%%%%%%%%%%%%%%%%%%%%%%

\begin{figure}[t]
    \centering
    \vs{-8mm}
    \includegraphics[width = 0.67\linewidth]{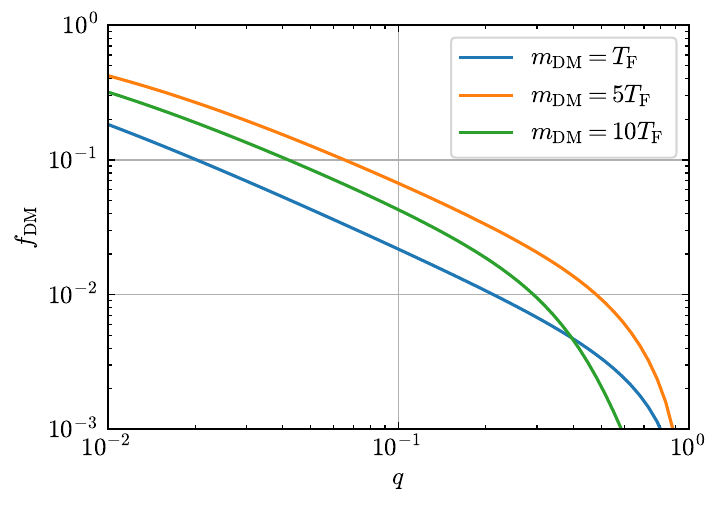}
    \caption{ Fraction of particle  DM abundance arising in a long MB phase from the first SC evaporation stage as a function of $q$ assuming $T_{F}\sim m_{\rm DM}$, i.e. around the limit for PBH to give rise to NCDM from evaporation and where we can expect the largest contribution of NCDM to the DM abundance. }
    \label{fig:param-spaceSC}
\end{figure}

Here we briefly discuss relatively long MB phases with $t_{\rm mb}> t_u$ where $t_u\sim 10^{17}$\,s denotes the age of the Universe. As visible in Eq.~(\ref{eq:OmBHlongMB}), this scenario is only valid for very suppressed initial PBH abundances. In the latter case, one can derive the maximal particle DM abundance that would arise from the SC phase of evaporation by fixing $\beta$ so as to account for all of the DM between particles and PBHs in the form of CDM. In order to obtain the maximal DM contribution, we set $m_{\rm DM }/T_{\Frm} \sim {\cal O} (1)$ so as to saturate the limit where it can be considered as NCDM. In this framework, the particle DM fraction from the first SC stage as a function of $q$ is shown in Fig.~\ref{fig:param-spaceSC} for three different choices of $m_{\rm DM }/T_{\Frm}= 1$ (blue curve), 5 (orange curve) and 10 (green curve). In the latter case, the DM mass becomes large enough compared to the BH temperature to suppress its abundance. The ratio of $m_{\rm DM }/T_{\Frm}= 5$ gives the maximal DM abundance while $m_{\rm DM }/T_{\Frm}= 1$ abundance is slightly smaller.

We see that a non-negligible particle DM abundance can be obtained for a MB phase that sets in at a very early stage of SC evaporation, i.e.~for  suppressed values of $q$. Even though $q\ll 1$ is not favoured by the arguments of Ref.~\cite{Dvali:2018xpy}, note that such a scenario would give rise to a dominating population of CDM in the form of relic PBHs in a long MB phase with a suppressed particle DM population that could behave as NCDM. Out of academic interest, this could, in particular, address the tension discussed in Ref.~\cite{Rogers:2023upm}, and be alleviated for CDM plus a percent fraction of a NCDM component.

%%%%%%%%%%%%%%%%%%%%%%%%%%%%%%%%%%%%%%%
\subsection{Short Memory-Burden Phase}
\label{sec:shortMB}
%%%%%%%%%%%%%%%%%%%%%%%%%%%%%%%%%%%%%%%

\begin{figure}[t]
    \centering
    \includegraphics[width = \linewidth]{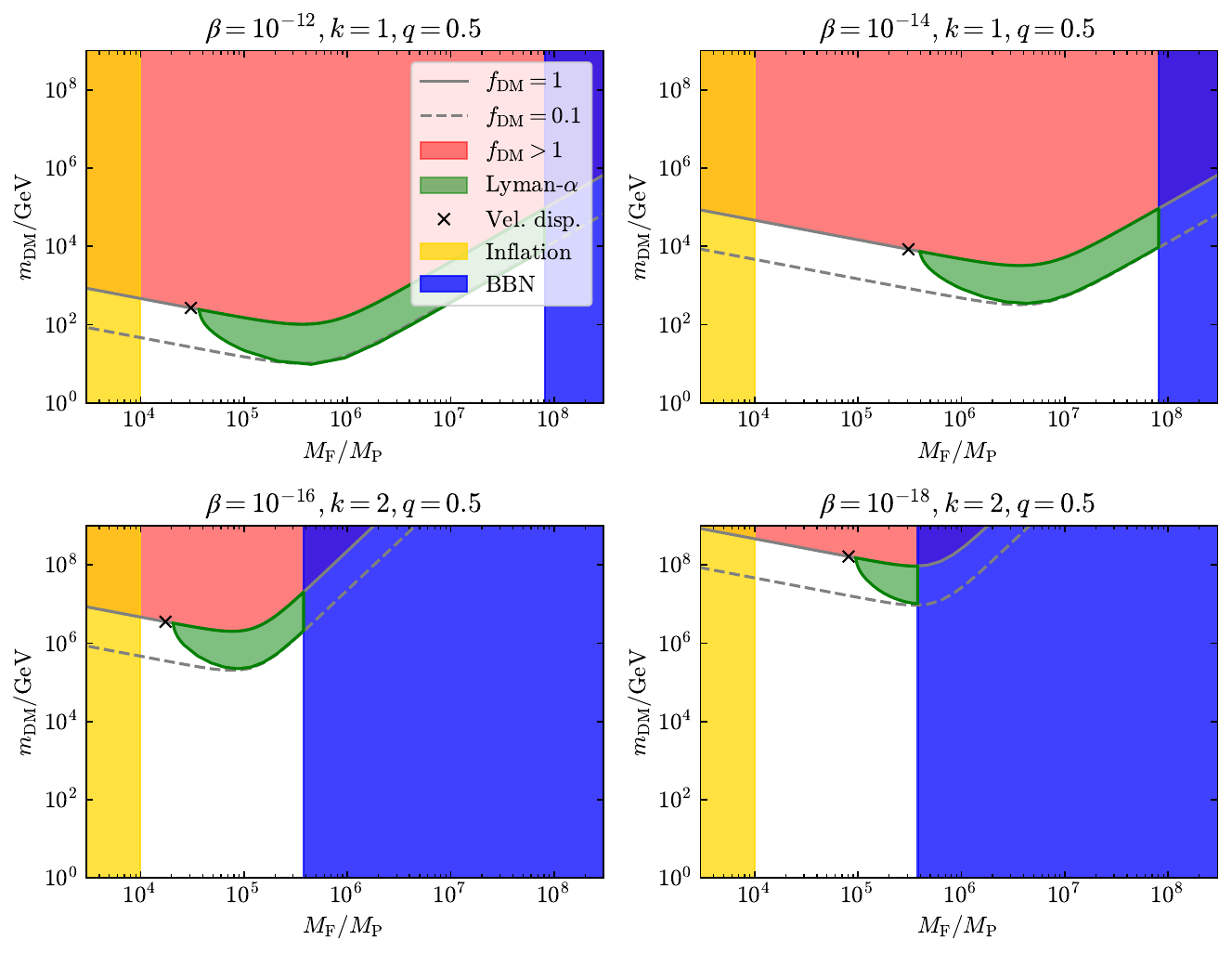}
    \caption{Allowed parameter space for DM production for $k = 1$ (top) and $k = 2$ (bottom). The left and right column show the results for different values of $\beta$. $q = 0.5$ is assumed in all cases. Bounds from BBN and inflation are displayed as the blue/yellow shaded region and they limit the range of allowed PBH masses. The gray lines highlight fractional DM abundances of $f_{\rm DM} = 1$ (continuous line) and $f_{\rm DM} = 0.1$ (dashed line), with the red shaded region corresponding to an overabundance of DM $f_{\rm DM}>1$. The green region is excluded due to Lyman-$\alpha$ constraints, computed using the area criterion. For $f_{\rm DM} = 1$ we indicate the estimate for the Lyman-$\alpha$ constraint based on Eq.~(\ref{eq:Kamada-mixBH}) as a black cross. }
    \label{fig:mdmvsMF}
\end{figure}

We now move to shorter MB phases with $t_{\rm mb}< t_u$, for which the NCDM imprint studied here can affect the viable parameter space. In Fig.~\ref{fig:mdmvsMF}, we show the viable parameter space in the $(M_{\Frm},\,m_{\rm DM})$ plane for $k = 1$ and $k = 2$ and two different values of the PBH abundance $\beta$. Note that this plot does not change qualitatively with $q$ except for very low values, which are not favored if theoretical arguments are considered. The shaded yellow region at low masses is excluded by CMB constraints on inflation~\cite{Akrami:2018odb}. 
\begin{comment}
     The latter places an upper bound on the Hubble rate at the end of inflation of $\sim 10^{14}$ GeV, translating into a lower bound on the PBH masses at formation $M_{\Frm}> 10^{4} M_{\rm P}$  directly related to the horizon mass at formation, see Sec.~\ref{sec:DMabund}. 
\end{comment}
On the other hand, BBN sets an upper bound on the PBH masses, displayed as a blue region. The latter becomes increasingly severe with increasing $k$ values, leaving no viable parameter space for $k\gtrsim4$ as visible in Fig.~\ref{fig:lifetimes} in the short MB region.

The correct relic abundance of particle DM, $\Omega_{\rm DM}\.h^{2} = 0.12$, is obtained along the continuous gray line of Fig.~\ref{fig:mdmvsMF}, while the dashed gray line indicates a smaller DM fraction $f_{\rm DM} = \Omega_{\rm DM}\.h^{2}/0.12$ of $10\%$. The change in slope of the gray lines for intermediate PBH masses indicates the transition between $\beta < \beta_{\rm c}$, for low PBH masses, and $\beta > \beta_{\rm c}$ at larger masses. In the case of RD ($\beta < \beta_{\rm c}$), the DM abundance increases with the PBH mass (for fixed DM particle mass) as opposed to the case of PBH domination ($\beta>\beta_{\rm c}$), see~Eq.(\ref{eq:OmDMshortmb}). The transition moves to lower $M_{\Frm}$ values when increasing $\beta$ as expected from Eq.~(\ref{eq:betacmb}), similar to the SC case as visible in Eq.~(\ref{eq:betacsc}), see also Ref.~\cite{Baldes:2020nuv}. The gray lines correspond to the analytic results provided in Sec.~(\ref{OmshortMB}) away from the smooth $\beta = \beta_{\rm c}$ transition that has been obtained numerically.\footnote{ Note that in these plots we have neglected the change in number of relativistic degrees of freedom in the early Universe that is expected to induce a small ($\lesssim$30\%) change in the gray curve in RD era.}

\begin{comment}
Also note that a small inflection in the gray curves appears in the $\beta> \beta_{\rm c}$ region, e.g.~at $M_{\Frm}\sim 10^{7} M_{\rm P}$ for $k = 1$ and $\beta =  10^{-12}$. This is due to a change in the number of relativistic degrees of freedom in the early Universe that affects the scale factor at evaporation $a_{\rm ev}$ which enters the computation of the DM relic abundance in a PBH dominated era. 
\end{comment}
The red region above the continuous gray lines is excluded as particle DM overcloses the Universe. 
 Note that increasing $k$ leads to slower evaporation in MB phase implying that a larger value of the DM mass is necessary to account for all the DM at fixed $M_{\Frm}$. 

The green regions in Fig.~\ref{fig:mdmvsMF} are excluded by the Lyman-$\alpha$ bounds derived in our work, making use of the area criterion introduced in Sec.~\ref{sec:areacrit}. The latter allows us to probe fractional NCDM contributions, setting the form of the region contours at low $M_{\Frm}$. We note that the area criterion becomes insensitive to NCDM components that make up less than 10\% of the total DM, where Lyman-$\alpha$ studies become less constraining, see the discussion in App.~\ref{sec:WCDM}. This determines the lower limit of the green area. Other probes on larger scales could conceivably extend this constraint. We also show with a black cross, the lower bound on the DM mass that one would derive making use of the velocity-dispersion estimate introduced in Eq.~(\ref{eq:Kamada-mixBH}) when NCDM from PBH evaporation accounts for all of the DM. The latter appear to be in good agreement with the results obtained by the area criterion. In the RD era ($\beta < \beta_{\rm c}$), the NCDM bound becomes weaker and allows for 100\% of the DM for $\beta \approx 0.015\.\beta_{\rm c}$. Notably, this is similar to the case of the SC treatment of PBH evaporation, see Ref~\cite{Decant:2021mhj, Auffinger:2020afu}. 

\begin{figure}[t]
    \centering
    \includegraphics[width = \linewidth]{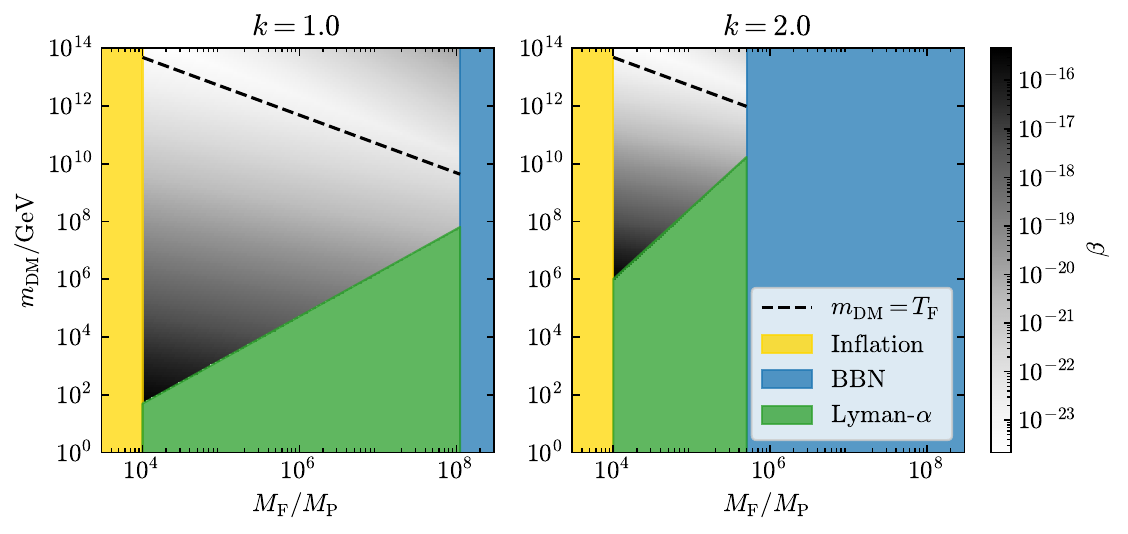}
    \caption{Parameter space for $q = 0.5$ as well as $k = 1$ (left) and $k = 2$ (right), for which all the DM can be produced by evaporating PBHs. The greyscale indicates the required value of $\beta$ in order to achieve the correct abundance. The colored shaded regions indicate the bounds from inflation (yellow), BBN (blue) and Lyman-$\alpha$ (green, computed using the area criterion). A change in behaviour happens at $m_{\rm DM}\sim T_{\rm F}$, indicated with a dashed black line.\\}
    \label{fig:mdmvsMFbetavar}
\end{figure}

Combining the different constraints, we see in Fig.~\ref{fig:mdmvsMF} that the viable parameter space for particle DM shrinks for larger values of $\beta$ and $k$, pushing the NCDM limit closer to the inflation bound. The region between the constraints from inflation and BBN also shrinks with increasing $k$ as the BBN bound extends to lower PBH masses. The dependency of the viable parameter space with $\beta$ is shown for $q = 0.5$ and two values of $k = 1$ and $k = 2$ in Fig.~\ref{fig:mdmvsMFbetavar} in the same $(M_{\Frm},m_{\rm DM})$ plane. The limits at low and large $M_{\Frm}$ are still set by inflation and BBN. The gradient of gray at intermediate masses indicates the value of $\beta$ that would give rise to all of the DM. Above the dashed black line, indicating $T_{\Frm} = m_{\rm DM}$, all the DM from PBH evaporation behaves as CDM. Fixed $\beta$ values would correspond to lines parallel to $T_{\Frm} = m_{\rm DM}$ dashed line. The lower limit of the gradient colored area is set by the NCDM constraint derived here. We see that the larger $\beta$ is, the smaller the viable region is.

\begin{figure}[h]
    \centering
    \includegraphics[width = 0.67\linewidth]{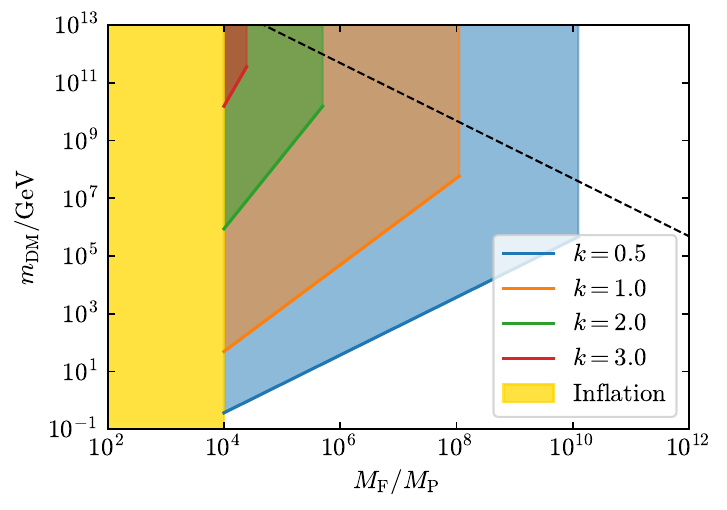}
    \caption{Allowed parameter space for various values of $k$ and $q = 0.5$. The yellow shaded region is excluded by CMB bounds on inflation. The other shaded colored areas are allowed for the $k$ values given in the legend. Those areas are bounded from below by the Lyman-$\alpha$ constraints derived here and from the right by BBN. }
    \label{fig:mdmvsMFkvar}
\end{figure}

The dependence of the viable parameter space on $k$ is shown for $q = 0.5$ with colored regions between $k = 0.5$ (blue area) and $k = 2$ (red area) in Fig.~\ref{fig:mdmvsMFkvar} in the same $(M_{\Frm},\,m_{\rm DM})$ plane. In each point of the viable parameter space, the value of $\beta$ is chosen in a way to give rise to the right relic abundance as illustrated in Fig.~\ref{fig:mdmvsMFbetavar}. As already mentioned above, the BBN bound moves to lower $M_{\Frm}$ values with increasing values of $k$ while the inflation bound is $M_{\Frm}$ independent. Furthermore, the NCDM bound moves to higher values of $m_{\rm DM}$ as expected from the MB contribution to our velocity estimate of Eq.~(\ref{eq:Kamada-mixBH}). This is due to longer PBH lifetimes at fixed $M_{\Frm}$ values giving rise to DM particles with higher velocity dispersion today.

%%%%%%%%%%%%%%%%%%%%%%%%%%%%%%%%%%%%%%%
\section{Summary and Conclusions}
\label{sec:concl}
%%%%%%%%%%%%%%%%%%%%%%%%%%%%%%%%%%%%%%%
In this paper, we have explored the implications of the memory-burden (MB) effect on the evaporation of primordial black holes (PBHs) and investigated in more detail its consequences for non-cold dark matter (NCDM) production and constraints. After re-deriving the DM abundance in a two-stage evaporation process involving a semi-classical (SC) and a MB phase, we obtained the NCDM velocity distribution and mean (squared) momenta for pure SC and MB phases. We have shown in particular that PBHs can emit DM particles with non-trivial phase-space distributions, which can result in multimodal velocity profiles in a small region of the parameter space. More generally, the velocity distribution deviates from a thermal profile, with its exact shape modified relative to the SC case due to the introduction of MB effects. For the purpose of our analysis, we made use of numerical simulations with \texttt{BlackHawk} to extract exact particle spectra from PBH evaporation, and when relevant, we provided matching analytical estimates.

Typically, the introduction of MB effects induces two DM particle populations from PBH evaporation with different velocity dispersions. In most of the relevant parameter space, one NCDM population arises from a recent MB-dominated evaporation phase and accounts for a fraction scaling as $\propto q^{2}$ of the total number of DM particles. This fraction is determined by the parameter $q$, which sets the PBH mass fraction at which the MB phase begins. The other DM population, originating from an earlier SC stage, behaves as CDM, its root-mean-squared velocity having been redshifted away. Using both analytical velocity-dispersion estimates and numerical simulations to evaluate the so-called area criterion introduced in Ref.~\cite{Murgia:2017lwo}, we reinterpreted existing Lyman-$\alpha$ constraints for thermal warm dark matter (WDM) to account for the presence of CDM + NCDM from PBH evaporation. This required implementing the DM velocity distribution in the Boltzmann code \texttt{CLASS}. Our analysis shows that velocity-dispersion estimates lead to more stringent constraints than the area criterion, as previously observed in another NCDM scenario (see, e.g., Ref.~\cite{Decant:2021mhj}). We also argue that the area criterion provides conservative bounds on the WDM mass in the context of Warm + CDM (see Appendix~\ref{sec:WCDM}, where we compare with results in the literature based on dedicated hydrodynamical simulations) and in the case of DM from PBH evaporation. Furthermore, we briefly comment on the possibility that PBHs may not fully evaporate in the MB phase and may contribute to the CDM fraction. In such a case, the particle-DM abundance typically accounts for only a few percent of the total relic density.

We have mapped the viable regions in the PBH--DM mass plane, accounting for bounds from inflation and Big Bang Nucleosynthesis, and incorporating our reinterpreted Lyman-$\alpha$ constraints based on the area criterion. The MB effect still allows PBHs to produce NCDM candidates, and accounting for all of the DM remains possible when PBH evaporation takes place in a radiation-dominated era with $\beta \lesssim 0.015\, \beta_{\rm c}$, as in the SC case. The viable DM parameter space shrinks significantly when considering larger initial PBH abundances or larger $k$ parameters defining the entropy-power suppression of the PBH evaporation rate in the MB phase. This confirms results from previous studies (see, e.g., Ref.~\cite{Barman:2024iht}), albeit with a more detailed treatment of the Lyman-$\alpha$ forest bound, which can play a critical role in determining the boundaries of the viable parameter space.

\newpage

%%%%%%%%%%%%%%%%%%%%%%%%%%%%%%%%%%%%%%%
\appendix
%%%%%%%%%%%%%%%%%%%%%%%%%%%%%%%%%%%%%%%

%%%%%%%%%%%%%%%%%%%%%%%%%%%%%%%%%%%%%%%
\section{More Details on The Phase Space Distribution}
\label{app:PSD}
%\section{More details on MB}
%%%%%%%%%%%%%%%%%%%%%%%%%%%%%%%%%%%%%%%

\begin{figure}[t]
    \centering
    \includegraphics[width = 0.7\linewidth]{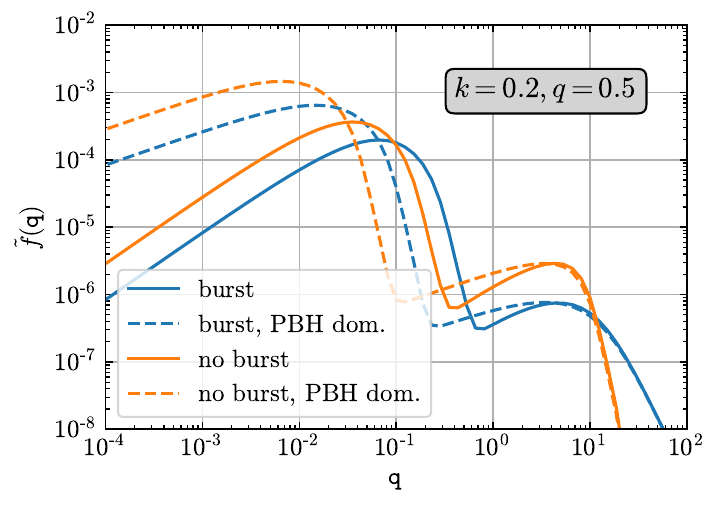}
    \caption{Rescaled momentum distributions $\tilde{f}(x)$ for $k = 0.2$ and $q = 0.5$. Same as in Fig.~\ref{fig:tidf2phases} but comparing RD to PBH domination (dashed curves).}
    \label{fig:tidf2phasesMD}
\end{figure}

In Fig.~\ref{fig:tidf2phasesMD} we show how the rescaled distribution for DM arising from PBH evaporation is affected by the fluid dominating the matter energy content of the Universe. In particular, for $q = 0.5$ and $k = 0.2$ the curves move from the continuous lines for production in a RD era to the dashed lines in a PBH dominated era. Indeed, when PBHs dominate, the scale-factor dependence on time changes, thus giving rise to more separated peaks and different slopes before the final tails.

%%%%%%%%%%%%%%%%%%%%%%%%%%%%%%%%%%%%%%%
\section{More Details on Lyman-$\alpha$ Constraints}
\label{sec:WCDM}
%%%%%%%%%%%%%%%%%%%%%%%%%%%%%%%%%%%%%%%

%%%%%%%%%%%%%%%%%%%%%%%%%%%%%%%%%%%%%%%
\subsection{Scales Probed by Lyman-$\alpha$ Forest}
%%%%%%%%%%%%%%%%%%%%%%%%%%%%%%%%%%%%%%%

As mentioned in Sec.~\ref{sec:WDM}, the flux power spectrum of distant quasars can constrain free-streaming properties of DM on the probed scales. In practice, the Lyman-$\alpha$ forest flux power spectrum is computed in terms of the dual of the velocity, which is a function of redshift and comoving position. The velocity wave number, ${\tt k}_v$ and the comoving wave number ${\tt k}$ are related through redshift dependent quantities: ${\tt k} = {\tt k}_v\.H(z)/(1+z)$, where $H$ denotes the Hubble rate, see, e.g., Refs.~\cite{Garzilli:2019qki} for a discussion. 

The datasets of XQ-100 and HIRES/MIKE used in Ref.~\cite{Irsic:2017ixq} to exclude $m_{\rm WDM}^{\text{Ly-}\alpha} = 5.3$ keV at 95\%CL for $f_{\rm WDM} = 1$, probed a redshift range between $z = $3 and 5.4, and velocity wave numbers between ${\tt k}_v = 0.003$ km/s and 0.08 km/s corresponding to a range of comoving wave numbers between ${\tt k}_{\rm min} = 0.3 h$/Mpc and ${\tt k}_{\rm max} =  10 h$/Mpc. In this paper, we use a slightly larger range of scales, reported in Eq.~(\ref{eq:krange}), following Ref.~\cite{Murgia:2017lwo} that proposed the area criterion as an estimate of the effect of NCDM on the flux power spectrum to translate WDM Lyman-$\alpha$ bounds to other NCDM scenarios, see Sec.~\ref{sec:areacrit} for details. We have checked that changing the considered range of scales to ${\tt k}\,\epsilon\,[0.3,\,10]\,h/$Mpc does not change significantly the range of excluded parameter space derived here from the moment that $\delta A_{\rm WDM}$ is properly recomputed.

%%%%%%%%%%%%%%%%%%%%%%%%%%%%%%%%%%%%%%%
\subsection{Warm and Cold Dark Matter as a Test}
%\subsection{WDM+CDM as a test}
%%%%%%%%%%%%%%%%%%%%%%%%%%%%%%%%%%%%%%%

\begin{figure}[t]
    \centering
    \includegraphics[width = 0.49\linewidth]{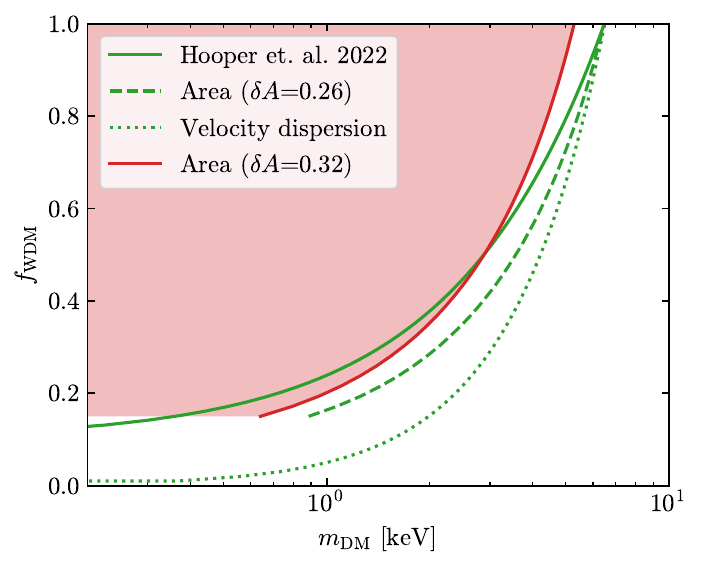}
    \includegraphics[width = 0.49\linewidth]{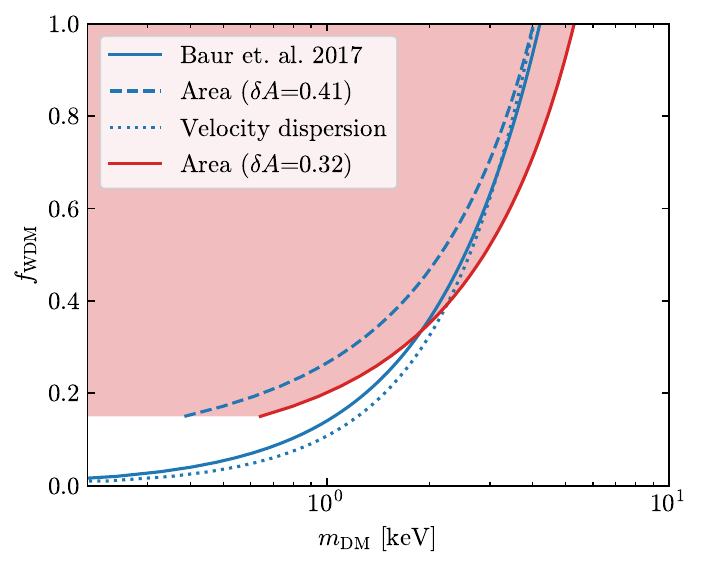}
        \caption{Comparison of the Lyman-$\alpha$ bound derived in Refs.~\cite{Baur:2017stq, Hooper:2022byl} for W+CDM to the ones described in Secs.~\ref{sec:warmcrit} and~\ref{sec:areacrit}. The left figure shows a comparison to the bound at 95\%CL from Ref.~\cite{Hooper:2022byl}, whereas the 2$\sigma$ bound from \cite{Baur:2017stq} is shown on the right. Their results are compared to the velocity dispersion estimate (dotted line) and the area criterion (dashed line) matched to \cite{Baur:2017stq,Hooper:2022byl} for $f_{\rm WDM} = 1$. In addition, we show as a red line the results from the area criterion for $m_{\rm WDM}^{\text{Ly-}\alpha} = 5.3$ keV at $f_{\rm WDM} = 1$ (in agreement with the findings of Ref.~\cite{Irsic:2017ixq}), that are being used in this work.}
    \label{fig:WDM}
\end{figure}

In the literature, constraints on CDM plus a fraction of thermal WDM have been carefully derived in the context of Lyman-$\alpha$ forest observations. In Fig.~\ref{fig:WDM}, displaying the WDM fraction, $f_{\rm WDM}$, we report the results of the analysis of Ref.~\cite{Hooper:2022byl, Baur:2017stq} as a function of the WDM mass $ m_{\rm WDM}$ with continuous lines, the area above those curves would be excluded at 95\%CL. In particular, by making use of a dedicated series of hydrodynamical simulations, and 
\begin{itemize}
    \item combining BOSS DR9 (see Ref.~\cite{BOSS:2013rpr}), XQ-100 (see Ref.~\cite{2016A&A...594A..91L}) and HIRES/MIKE  (compiled in Ref.~\cite{Viel:2013fqw}) data sets, the 2-$\sigma$ bound was shown to follow
    \begin{equation}
       f_{\rm WDM}\lesssim 0.14\,({\rm keV}/ m_{\rm WDM})^{-1.37}\quad {\rm [Baur\; 2017]}
    \label{eq:mflyaBaur}
    \end{equation}
    in Ref.~\cite{Baur:2017stq}. The bound from Eq.~\ref{eq:mflyaBaur} is plotted with a blue continuous line in the right plot of Fig.~\ref{fig:WDM} and referred to as Baur 2017. Note that Eq.~(\ref{eq:mflyaBaur}) excludes a WDM candidate with $m_{\rm WDM}^{\text{Ly-}\alpha} = 4.2\.$keV for $f_{\rm WDM} = 1$. Note that the fit of the 2-$\sigma$ bound of Eq.~(\ref{eq:mflyaBaur}) was obtained by interpolating a grid of simulations considering $m_{\rm WDM} > $ 0.7\.keV. This fit corresponds to an extrapolation for $f_{\rm WDM}\lesssim 0.1$.

    \item using MIKE/HIRES data, a more recent analysis fitted their 95\% CL bound to
    \begin{equation}
    m_{\rm WDM}\gtrsim 7.2\,{\rm keV}\,(f_{\rm WDM}-0.1)\quad {\rm [Hooper\;2022]}
    \label{eq:mflyaHooper}
    \end{equation}
    in Ref.~\cite{Hooper:2022byl}. This bound is shown with a green continuous line in the right plot of Fig.~\ref{fig:WDM} and is referred to as Hooper 2022. In particular, those results exclude a WDM candidate with $m_{\rm WDM}^{\text{Ly-}\alpha} = $6.5 keV for $f_{\rm WDM} = 1$. Note that the Hooper 2022 bound is stronger than the Baur 2017 bound for large fractions of WDM while it is less stringent for $f_{\rm WDM}<0.4$.
\end{itemize}
The dependence of the bound in the $(m_{\rm WDM}, f_{\rm WDM})$ plane is quite different between the two analyses. However, note that Ref.~\cite{Garcia-Gallego:2025kiw} more recently provided an updated analysis, and their 95\%CL constraints (including corrections) are in very good agreement with the 95\%CL bound of Hooper 2022. In any case, we cannot directly use these results for the NCDM scenarios discussed in Sec.~\ref{sec:twopop}. Indeed, even in the Case~\ref{lab:CWDM} of CDM plus a fraction of NCDM, the NCDM distribution differs from a Fermi-Dirac one, which is the basis for the derivation of the bounds reported in Eqs.~(\ref{eq:mflyaBaur}) and~(\ref{eq:mflyaHooper}). 

Here, we test the methodologies presented in Secs.~\ref{sec:warmcrit} and~\ref{sec:areacrit} against W+CDM scenarios. Our recasted bounds on W+CDM scenarios are reported in both plots of Fig.~\ref{fig:WDM} with dotted lines when considering the velocity dispersion argument, while the dashed lines were obtained using the area criterion. In particular, for the velocity dispersion constraints, Eq.~(\ref{eq:Kamada-mix}) translates into a bound
\begin{equation}
 f_{\rm WDM}\lesssim ( m_{\rm WDM}/{\rm keV})^{8/5}\times \Big( m_{\rm WDM}^{\text{Ly-}\alpha}/{\rm keV} \Big)^{-8/5}\,,
  \label{eq:Kamada-mixWCDM}
\end{equation}
where we have used ${\tt T_{\rm NCDM}} = 0.16\,(f_{\rm WDM} ({\rm keV}/ m_{\rm WDM}))^{1/3}$ for the WDM component (while ${\tt T_{\rm NCDM}} = 0$ for CDM) as well as $m_{\rm WDM}^{\text{Ly-}\alpha} =  4.2$ keV and 6.5 keV for the dotted blue and green lines in the right and left plots, respectively. In particular, we see that Eq.~(\ref{eq:Kamada-mixWCDM}) reproduces relatively well the dependence of the fit on the 2-$\sigma$ bound derived by Baur {\it et al.} in 2017. The continuous and dotted blue lines are indeed very much alike. On the other hand, Eq.~(\ref{eq:Kamada-mixWCDM}) definitively overestimates the bound of Hooper {\it et al.}~especially at low WDM fractions when assuming the same excluded WDM mass at $f_{\rm WDM} = 1$.

In order to extract the area criterion constraints, we have computed the relative area values $\delta A_{\rm WDM}$ for $m_{\rm WDM}^{\text{Ly-}\alpha} =  4.2$ keV and 6.5 keV considering $f_{\rm WDM} =  1$ and we have obtained $\delta A_{\rm WDM} = 0.41$ and 0.26. Larger $\delta A_{\rm WDM}$ corresponding to a less stringent WDM bound (i.e.~excluding smaller WDM masses or equivalently stronger suppressions of the matter power spectrum at small scales). The resulting estimates of the limit, obtained by varying the WDM fraction and simulating the corresponding effect on the matter power spectrum with {\tt CLASS}, are shown with dashed blue and green lines in the plots of Fig.~\ref{fig:WDM}. As we can see, the area criterion estimate of the bounds provides more conservative constraints than the corresponding velocity dispersion estimate (dotted lines).

In this paper, we consider the lower WDM bound of $m_{\rm WDM}^{\text{Ly-}\alpha} = 5.3$ keV for $f_{\rm WDM} = 1$ for which $\delta A_{\rm WDM} = 0.32$ as reported in Sec.~\ref{sec:areacrit}, following~\cite{Murgia:2017lwo}. The latter is projected onto the $(m_{\rm WDM}, f_{\rm WDM})$ plane as the dashed red line, and the corresponding estimate of the excluded parameter space is shaded in red. We can see that it provides a good compromise between Hooper 2022, which is more constraining at large WDM fractions, and Baur 2017, which is more constraining at low WDM fractions. As a result, in this paper, we make use of the area criterion saturating the $m_{\rm WDM}^{\text{Ly-}\alpha} = 5.3$ keV bound for $f_{\rm WDM} = 1$ as a conservative estimate of Lyman-$\alpha$ constraints, at least with respect to the velocity dispersion estimate of Eq.~(\ref{eq:Kamada-mix}).

%%%%%%%%%%%%%%%%%%%%%%%%%%%%%%%%%%%%%%%
\acknowledgments

VT is supported by the Excellence Cluster ORIGINS, which is funded by the Deutsche Forschungsgemeinschaft (DFG, German Research Foundation) under Germany’s Excellence Strategy - EXC-2094 - 390783311. FK acknowledges support from and hospitality of the Max Planck Institute for Physics (MPP). LLH is supported by the Fonds de la Recherche Scientifique F.R.S.-FNRS through a senior research associate (LLH) and acknowledges the support of the FNRS research grant number J.0134.24, the ARC program of the Federation Wallonie-Bruxelles and the IISN convention No. 4.4503.15. LLH and MH are also member of BLU-ULB (Brussels Laboratory of the Universe, blu.ulb.be).
 
\begin{comment}

%%%%%%%%%%%%%%%%%%%%%%%%%%%%%%%%%%%%%%%
\paragraph{Note added.} This is also a good position for notes added
after the paper has been written.
\end{comment}

%%%%%%%%%%%%%%%%%%%%%%%%%%%%%%%%%%%%%%%
% Bibliography

%% [A] Recommended: using JHEP.bst file
\bibliographystyle{JHEP}
\bibliography{biblio.bib}

@article{Carr:2016hva,
    author = "Carr, B. J. and Kohri, Kazunori and Sendouda, Yuuiti and Yokoyama, Jun'ichi",
    title = "{Constraints on primordial black holes from the Galactic gamma-ray background}",
    eprint = "1604.05349",
    archivePrefix = "arXiv",
    primaryClass = "astro-ph.CO",
    reportNumber = "RESCEU-16-16, KEK-TH-1895, KEK-COSMO-193",
    doi = "10.1103/PhysRevD.94.044029",
    journal = "Phys. Rev. D",
    volume = "94",
    number = "4",
    pages = "044029",
    year = "2016"
}

@article{Dvali:2025sog,
    author = "Dvali, Gia",
    title = "{Swift Memory Burden in Merging Black Holes: how information load affects black hole's classical dynamics}",
    eprint = "2509.22540",
    archivePrefix = "arXiv",
    primaryClass = "hep-th",
    month = "9",
    year = "2025"
}

@article{Keith:2020jww,
    author = "Keith, Celeste and Hooper, Dan and Blinov, Nikita and McDermott, Samuel D.",
    title = "{Constraints on Primordial Black Holes From Big Bang Nucleosynthesis Revisited}",
    eprint = "2006.03608",
    archivePrefix = "arXiv",
    primaryClass = "astro-ph.CO",
    reportNumber = "FERMILAB-PUB-20-224-A",
    doi = "10.1103/PhysRevD.102.103512",
    journal = "Phys. Rev. D",
    volume = "102",
    number = "10",
    pages = "103512",
    year = "2020"
}

@article{Miyama:1978mp,
    author = "Miyama, Shoken and Sato, Katsuhiko",
    title = "{The Upper Bound of the Number Density of Primordial Black Holes From the Big Bang Nucleosynthesis}",
    reportNumber = "KUNS 426",
    doi = "10.1143/PTP.59.1012",
    journal = "Prog. Theor. Phys.",
    volume = "59",
    pages = "1012",
    year = "1978"
}

@article{Ivanov:2025pbu,
    author = "Ivanov, Mikhail M. and Trifinopoulos, Sokratis",
    title = "{Effective Field Theory Constraints on Primordial Black Holes from the High-Redshift Lyman-$\alpha$ Forest}",
    eprint = "2508.04767",
    archivePrefix = "arXiv",
    primaryClass = "astro-ph.CO",
    reportNumber = "MIT-CTP/5895, CERN-TH-2025-155",
    month = "8",
    year = "2025"
}

@article{Murgia:2019duy,
    author = "Murgia, Riccardo and Scelfo, Giulio and Viel, Matteo and Raccanelli, Alvise",
    title = "{Lyman-{\ensuremath{\alpha}} Forest Constraints on Primordial Black Holes as Dark Matter}",
    eprint = "1903.10509",
    archivePrefix = "arXiv",
    primaryClass = "astro-ph.CO",
    reportNumber = "CERN-TH-2019-029",
    doi = "10.1103/PhysRevLett.123.071102",
    journal = "Phys. Rev. Lett.",
    volume = "123",
    number = "7",
    pages = "071102",
    year = "2019"
}

@article{Sun:2025ksr,
    author = "Sun, Yitian and Foster, Joshua W. and Mu{\~n}oz, Julian B.",
    title = "{Constraining inhomogeneous energy injection from annihilating dark matter and primordial black holes with 21-cm cosmology}",
    eprint = "2509.22772",
    archivePrefix = "arXiv",
    primaryClass = "hep-ph",
    month = "9",
    year = "2025"
}

@article{Clark:2018ghm,
    author = "Clark, Steven and Dutta, Bhaskar and Gao, Yu and Ma, Yin-Zhe and Strigari, Louis E.",
    title = "{21 cm limits on decaying dark matter and primordial black holes}",
    eprint = "1803.09390",
    archivePrefix = "arXiv",
    primaryClass = "astro-ph.HE",
    reportNumber = "MI-TH-1879",
    doi = "10.1103/PhysRevD.98.043006",
    journal = "Phys. Rev. D",
    volume = "98",
    number = "4",
    pages = "043006",
    year = "2018"
}

@article{Tashiro:2012qe,
    author = "Tashiro, Hiroyuki and Sugiyama, Naoshi",
    title = "{The effect of primordial black holes on 21 cm fluctuations}",
    eprint = "1207.6405",
    archivePrefix = "arXiv",
    primaryClass = "astro-ph.CO",
    doi = "10.1093/mnras/stt1493",
    journal = "Mon. Not. Roy. Astron. Soc.",
    volume = "435",
    pages = "3001",
    year = "2013"
}

@article{Zantedeschi:2024ram,
    author = "Zantedeschi, Michael and Visinelli, Luca",
    title = "{Ultralight black holes as sources of high-energy particles}",
    eprint = "2410.07037",
    archivePrefix = "arXiv",
    primaryClass = "astro-ph.HE",
    doi = "10.1016/j.dark.2025.102034",
    journal = "Phys. Dark Univ.",
    volume = "49",
    pages = "102034",
    year = "2025"
}

@article{Hawking:1971ei,
    author = "Hawking, Stephen",
    title = "{Gravitationally collapsed objects of very low mass}",
    doi = "10.1093/mnras/152.1.75",
    journal = "Mon. Not. Roy. Astron. Soc.",
    volume = "152",
    pages = "75",
    year = "1971"
}

@article{Chaudhuri:2025asm,
    author = "Chaudhuri, Arnab and Kohri, Kazunori and Thoss, Valentin",
    title = "{New bounds on Memory Burdened Primordial Black Holes from Big Bang Nucleosynthesis}",
    eprint = "2506.20717",
    archivePrefix = "arXiv",
    primaryClass = "astro-ph.CO",
    reportNumber = "KEK-TH-2735, KEK-Cosmo-0384",
    month = "6",
    year = "2025"
}

@article{Chianese:2024rsn,
    author = "Chianese, Marco and Boccia, Andrea and Iocco, Fabio and Miele, Gennaro and Saviano, Ninetta",
    title = "{Light burden of memory: Constraining primordial black holes with high-energy neutrinos}",
    eprint = "2410.07604",
    archivePrefix = "arXiv",
    primaryClass = "astro-ph.HE",
    doi = "10.1103/PhysRevD.111.063036",
    journal = "Phys. Rev. D",
    volume = "111",
    number = "6",
    pages = "063036",
    year = "2025"
}

@article{Kohri:2024qpd,
    author = "Kohri, Kazunori and Terada, Takahiro and Yanagida, Tsutomu T.",
    title = "{Induced gravitational waves probing primordial black hole dark matter with the memory burden effect}",
    eprint = "2409.06365",
    archivePrefix = "arXiv",
    primaryClass = "astro-ph.CO",
    reportNumber = "KEK-TH-2654, KEK-Cosmo-0358",
    doi = "10.1103/PhysRevD.111.063543",
    journal = "Phys. Rev. D",
    volume = "111",
    number = "6",
    pages = "063543",
    year = "2025"
}

@article{Bhaumik:2024qzd,
    author = "Bhaumik, Nilanjandev and Haque, Md Riajul and Jain, Rajeev Kumar and Lewicki, Marek",
    title = "{Memory burden effect mimics reheating signatures on SGWB from ultra-low mass PBH domination}",
    eprint = "2409.04436",
    archivePrefix = "arXiv",
    primaryClass = "astro-ph.CO",
    doi = "10.1007/JHEP10(2024)142",
    journal = "JHEP",
    volume = "10",
    pages = "142",
    year = "2024"
}

@article{Dvali:2018xpy,
    author = "Dvali, Gia",
    title = "{A Microscopic Model of Holography: Survival by the Burden of Memory}",
    eprint = "1810.02336",
    archivePrefix = "arXiv",
    primaryClass = "hep-th",
    month = "10",
    year = "2018"
}

@article{Carr:2020xqk,
    author = "Carr, Bernard and Kuhnel, Florian",
    title = "{Primordial Black Holes as Dark Matter: Recent Developments}",
    eprint = "2006.02838",
    archivePrefix = "arXiv",
    primaryClass = "astro-ph.CO",
    doi = "10.1146/annurev-nucl-050520-125911",
    journal = "Ann. Rev. Nucl. Part. Sci.",
    volume = "70",
    pages = "355--394",
    year = "2020"
}

@article{Escriva:2022duf,
    author = "Escriv\`a, Albert and Kuhnel, Florian and Tada, Yuichiro",
    title = "{Primordial Black Holes}",
    eprint = "2211.05767",
    archivePrefix = "arXiv",
    primaryClass = "astro-ph.CO",
    doi = "10.1016/B978-0-32-395636-9.00012-8",
    month = "11",
    year = "2022"
}

@article{Dvali:2021byy,
    author = {Dvali, Gia and K\"uhnel, Florian and Zantedeschi, Michael},
    title = "{Primordial black holes from confinement}",
    eprint = "2108.09471",
    archivePrefix = "arXiv",
    primaryClass = "hep-ph",
    doi = "10.1103/PhysRevD.104.123507",
    journal = "Phys. Rev. D",
    volume = "104",
    number = "12",
    pages = "123507",
    year = "2021"
}

@article{Alexandre:2024nuo,
    author = "Alexandre, Ana and Dvali, Gia and Koutsangelas, Emmanouil",
    title = "{New mass window for primordial black holes as dark matter from the memory burden effect}",
    eprint = "2402.14069",
    archivePrefix = "arXiv",
    primaryClass = "hep-ph",
    doi = "10.1103/PhysRevD.110.036004",
    journal = "Phys. Rev. D",
    volume = "110",
    number = "3",
    pages = "036004",
    year = "2024"
}

@article{Dvali:2024hsb,
    author = "Dvali, Gia and Valbuena-Berm\'udez, Juan Sebasti\'an and Zantedeschi, Michael",
    title = "{Memory burden effect in black holes and solitons: Implications for PBH}",
    eprint = "2405.13117",
    archivePrefix = "arXiv",
    primaryClass = "hep-th",
    doi = "10.1103/PhysRevD.110.056029",
    journal = "Phys. Rev. D",
    volume = "110",
    number = "5",
    pages = "056029",
    year = "2024"
}

@article{Thoss:2024hsr,
    author = "Thoss, Valentin and Burkert, Andreas and Kohri, Kazunori",
    title = "{Breakdown of hawking evaporation opens new mass window for primordial black holes as dark matter candidate}",
    eprint = "2402.17823",
    archivePrefix = "arXiv",
    primaryClass = "astro-ph.CO",
    reportNumber = "KEK-TH-2605;KEK-Cosmo-0339;KEK-QUP-2024-0003",
    doi = "10.1093/mnras/stae1098",
    journal = "Mon. Not. Roy. Astron. Soc.",
    volume = "532",
    number = "1",
    pages = "451--459",
    year = "2024"
}

@ARTICLE{Hawking1975,
       author = {{Hawking}, S.~W.},
        title = "{Particle creation by black holes}",
      journal = {Communications in Mathematical Physics},
         year = 1975,
        month = aug,
       volume = {43},
       number = {3},
        pages = {199-220},
          doi = {10.1007/BF02345020},
       adsurl = {https://ui.adsabs.harvard.edu/abs/1975CMaPh..43..199H},
      related = {Hawking1975Erratum},
relatedstring = {Erratum:},
}

@article{Carr:2020gox,
    author = "Carr, Bernard and Kohri, Kazunori and Sendouda, Yuuiti and Yokoyama, Jun'ichi",
    archivePrefix = "arXiv",
    eprint = "2002.12778",
    month = "2",
    primaryClass = "astro-ph.CO",
    reportNumber = "RESCEU-03/20; KEK-Cosmo-249; KEK-TH-2199; IPMU20-0024",
    title = "{Constraints on Primordial Black Holes}",
    year = "2020"
}

@article{Carr:1974nx,
    author = "Carr, Bernard J. and Hawking, S.W.",
    journal = "Mon. Not. Roy. Astron. Soc.",
    pages = "399--415",
    title = "{Black holes in the early Universe}",
    volume = "168",
    year = "1974"
}

@article{Aghanim:2018eyx,
    author = "Aghanim, N. and others",
    archivePrefix = "arXiv",
    collaboration = "Planck",
    eprint = "1807.06209",
    month = "7",
    primaryClass = "astro-ph.CO",
    title = "{Planck 2018 results. VI. Cosmological parameters}",
    year = "2018"
}

@article{Akrami:2018odb,
    author = "Akrami, Y. and others",
    archivePrefix = "arXiv",
    collaboration = "Planck",
    eprint = "1807.06211",
    month = "7",
    primaryClass = "astro-ph.CO",
    title = "{Planck 2018 results. X. Constraints on inflation}",
    year = "2018"
}

@article{Hooper:2020evu,
      author         = "Hooper, Dan and Krnjaic, Gordan and March-Russell, John
                        and McDermott, Samuel D. and Petrossian-Byrne, Rudin",
      title          = "{Hot Gravitons and Gravitational Waves From Kerr Black
                        Holes in the Early Universe}",
      year           = "2020",
      eprint         = "2004.00618",
      archivePrefix  = "arXiv",
      primaryClass   = "astro-ph.CO",
      reportNumber   = "FERMILAB-PUB-20-125-A-T",
      SLACcitation   = "%%CITATION = ARXIV:2004.00618;%%"
}

@article{Murgia:2017lwo,
      author         = "Murgia, Riccardo and Merle, Alexander and Viel, Matteo
                        and Totzauer, Maximilian and Schneider, Aurel",
      title          = "{"Non-cold" dark matter at small scales: a general
                        approach}",
      journal        = "JCAP",
      volume         = "1711",
      year           = "2017",
      pages          = "046",
      doi            = "10.1088/1475-7516/2017/11/046",
      eprint         = "1704.07838",
      archivePrefix  = "arXiv",
      primaryClass   = "astro-ph.CO",
      SLACcitation   = "%%CITATION = ARXIV:1704.07838;%%"
}

@article{Lesgourgues:2011rh,
      author         = "Lesgourgues, Julien and Tram, Thomas",
      title          = "{The Cosmic Linear Anisotropy Solving System (CLASS) IV:
                        efficient implementation of non-cold relics}",
      journal        = "JCAP",
      volume         = "1109",
      year           = "2011",
      pages          = "032",
      doi            = "10.1088/1475-7516/2011/09/032",
      eprint         = "1104.2935",
      archivePrefix  = "arXiv",
      primaryClass   = "astro-ph.CO",
      reportNumber   = "CERN-PH-TH-2011-084, LAPTH-012-11",
      SLACcitation   = "%%CITATION = ARXIV:1104.2935;%%"
}

@article{Arbey:2019mbc,
      author         = "Arbey, Alexandre and Auffinger, Jérémy",
      title          = "{BlackHawk: A public code for calculating the Hawking
                        evaporation spectra of any black hole distribution}",
      year           = "2019",
      eprint         = "1905.04268",
      archivePrefix  = "arXiv",
      primaryClass   = "gr-qc",
      reportNumber   = "CERN-TH-2019-067",
      SLACcitation   = "%%CITATION = ARXIV:1905.04268;%%"
}

@article{Lennon:2017tqq,
      author         = "Lennon, Olivier and March-Russell, John and
                        Petrossian-Byrne, Rudin and Tillim, Hannah",
      title          = "{Black Hole Genesis of Dark Matter}",
      journal        = "JCAP",
      volume         = "1804",
      year           = "2018",
      number         = "04",
      pages          = "009",
      doi            = "10.1088/1475-7516/2018/04/009",
      eprint         = "1712.07664",
      archivePrefix  = "arXiv",
      primaryClass   = "hep-ph",
      SLACcitation   = "%%CITATION = ARXIV:1712.07664;%%"
}

@article{Baldes:2014rda,
      author         = "Baldes, Iason and Bell, Nicole F. and Millar, Alexander
                        and Petraki, Kalliopi and Volkas, Raymond R.",
      title          = "{The role of CP violating scatterings in baryogenesis -
                        case study of the neutron portal}",
      journal        = "JCAP",
      volume         = "1411",
      year           = "2014",
      number         = "11",
      pages          = "041",
      doi            = "10.1088/1475-7516/2014/11/041",
      eprint         = "1410.0108",
      archivePrefix  = "arXiv",
      primaryClass   = "hep-ph",
      SLACcitation   = "%%CITATION = ARXIV:1410.0108;%%"
}

@article{Carr:2016drx,
      author         = "Carr, Bernard and Kuhnel, Florian and Sandstad, Marit",
      title          = "{Primordial Black Holes as Dark Matter}",
      journal        = "Phys. Rev.",
      volume         = "D 94",
      year           = "2016",
      number         = "8",
      pages          = "083504",
      doi            = "10.1103/PhysRevD.94.083504",
      eprint         = "1607.06077",
      archivePrefix  = "arXiv",
      primaryClass   = "astro-ph.CO",
      reportNumber   = "NORDITA-2016-83",
      SLACcitation   = "%%CITATION = ARXIV:1607.06077;%%"
}

@article{Irsic:2017ixq,
      author         = "Iršič, Vid and others",
      title          = "{New Constraints on the free-streaming of warm dark
                        matter from intermediate and small scale Lyman-$\alpha$
                        forest data}",
      journal        = "Phys. Rev.",
      volume         = "D96",
      year           = "2017",
      number         = "2",
      pages          = "023522",
      doi            = "10.1103/PhysRevD.96.023522",
      eprint         = "1702.01764",
      archivePrefix  = "arXiv",
      primaryClass   = "astro-ph.CO",
      SLACcitation   = "%%CITATION = ARXIV:1702.01764;%%"
}

@article{Garzilli:2019qki,
    author = "Garzilli, A. and Ruchayskiy, O. and Magalich, A. and Boyarsky, A.",
    archivePrefix = "arXiv",
    eprint = "1912.09397",
    month = "12",
    primaryClass = "astro-ph.CO",
    title = "{How warm is too warm? Towards robust Lyman-$\alpha$ forest bounds on warm dark matter}",
    year = "2019"
}

@article{Baur:2017stq,
    author = "Baur, Julien and Palanque-Delabrouille, Nathalie and Yeche, Christophe and Boyarsky, Alexey and Ruchayskiy, Oleg and Armengaud, Eric and Lesgourgues, Julien",
    archivePrefix = "arXiv",
    doi = "10.1088/1475-7516/2017/12/013",
    eprint = "1706.03118",
    journal = "JCAP",
    pages = "013",
    primaryClass = "astro-ph.CO",
    title = "{Constraints from Ly-$\alpha$ forests on non-thermal dark matter including resonantly-produced sterile neutrinos}",
    volume = "12",
    year = "2017"
}

@article{Boyarsky:2008xj,
    author = "Boyarsky, Alexey and Lesgourgues, Julien and Ruchayskiy, Oleg and Viel, Matteo",
    archivePrefix = "arXiv",
    doi = "10.1088/1475-7516/2009/05/012",
    eprint = "0812.0010",
    journal = "JCAP",
    pages = "012",
    primaryClass = "astro-ph",
    reportNumber = "CERN-PH-TH-2008-234, LAPTH-1290-08",
    title = "{Lyman-alpha constraints on warm and on warm-plus-cold dark matter models}",
    volume = "05",
    year = "2009"
}

@article{Carr:1975qj,
	author = "Carr, Bernard J.",
	doi = "10.1086/153853",
	journal = "Astrophys. J.",
	pages = "1--19",
	title = "{The Primordial black hole mass spectrum}",
	volume = "201",
	year = "1975"
}

@article{Dvali:2020wft,
    author = "Dvali, Gia and Eisemann, Lukas and Michel, Marco and Zell, Sebastian",
    title = "{Black hole metamorphosis and stabilization by memory burden}",
    eprint = "2006.00011",
    archivePrefix = "arXiv",
    primaryClass = "hep-th",
    doi = "10.1103/PhysRevD.102.103523",
    journal = "Phys. Rev. D",
    volume = "102",
    number = "10",
    pages = "103523",
    year = "2020"
}

@article{Decant:2021mhj,
    author = "Decant, Quentin and Heisig, Jan and Hooper, Deanna C. and Lopez-Honorez, Laura",
    title = "{Lyman-\ensuremath{\alpha} constraints on freeze-in and superWIMPs}",
    eprint = "2111.09321",
    archivePrefix = "arXiv",
    primaryClass = "astro-ph.CO",
    reportNumber = "ULB-TH/21-20, TTK-21-46, HIP-2021-38/TH",
    doi = "10.1088/1475-7516/2022/03/041",
    journal = "JCAP",
    volume = "03",
    pages = "041",
    year = "2022"
}

@article{Ballesteros:2020adh,
    author = "Ballesteros, Guillermo and Garcia, Marcos A. G. and Pierre, Mathias",
    title = "{How warm are non-thermal relics? Lyman-$\alpha$ bounds on out-of-equilibrium dark matter}",
    eprint = "2011.13458",
    archivePrefix = "arXiv",
    primaryClass = "hep-ph",
    reportNumber = "IFT-UAM/CSIC-20-135",
    doi = "10.1088/1475-7516/2021/03/101",
    journal = "JCAP",
    volume = "03",
    pages = "101",
    year = "2021"
}

@article{Bae:2017dpt,
    author = "Bae, Kyu Jung and Kamada, Ayuki and Liew, Seng Pei and Yanagi, Keisuke",
    title = "{Light axinos from freeze-in: production processes, phase space distributions, and Ly-$\alpha$ forest constraints}",
    eprint = "1707.06418",
    archivePrefix = "arXiv",
    primaryClass = "hep-ph",
    reportNumber = "CTPU-17-25, UT-17-25",
    doi = "10.1088/1475-7516/2018/01/054",
    journal = "JCAP",
    volume = "01",
    pages = "054",
    year = "2018"
}

@article{Baldes:2020nuv,
    author = "Baldes, Iason and Decant, Quentin and Hooper, Deanna C. and Lopez-Honorez, Laura",
    title = "{Non-Cold Dark Matter from Primordial Black Hole Evaporation}",
    eprint = "2004.14773",
    archivePrefix = "arXiv",
    primaryClass = "astro-ph.CO",
    reportNumber = "ULB-TH/20-05",
    doi = "10.1088/1475-7516/2020/08/045",
    journal = "JCAP",
    volume = "08",
    pages = "045",
    year = "2020"
}

@article{Haque:2024eyh,
    author = "Haque, Md Riajul and Maity, Suvashis and Maity, Debaprasad and Mambrini, Yann",
    title = "{Quantum effects on the evaporation of PBHs: contributions to dark matter}",
    eprint = "2404.16815",
    archivePrefix = "arXiv",
    primaryClass = "hep-ph",
    doi = "10.1088/1475-7516/2024/07/002",
    journal = "JCAP",
    volume = "07",
    pages = "002",
    year = "2024"
}

@article{Cheek:2022mmy,
    author = "Cheek, Andrew and Heurtier, Lucien and Perez-Gonzalez, Yuber F. and Turner, Jessica",
    title = "{Evaporation of primordial black holes in the early Universe: Mass and spin distributions}",
    eprint = "2212.03878",
    archivePrefix = "arXiv",
    primaryClass = "hep-ph",
    reportNumber = "IPPP/22/79",
    doi = "10.1103/PhysRevD.108.015005",
    journal = "Phys. Rev. D",
    volume = "108",
    number = "1",
    pages = "015005",
    year = "2023"
}

@article{Barman:2024iht,
    author = "Barman, Basabendu and Haque, Md Riajul and Zapata, \'Oscar",
    title = "{Gravitational wave signatures of cogenesis from a burdened PBH}",
    eprint = "2405.15858",
    archivePrefix = "arXiv",
    primaryClass = "astro-ph.CO",
    doi = "10.1088/1475-7516/2024/09/020",
    journal = "JCAP",
    volume = "09",
    pages = "020",
    year = "2024"
}

@article{Viel:2013fqw,
    author = "Viel, Matteo and Becker, George D. and Bolton, James S. and Haehnelt, Martin G.",
    title = "{Warm dark matter as a solution to the small scale crisis: New constraints from high redshift Lyman-\ensuremath{\alpha} forest data}",
    eprint = "1306.2314",
    archivePrefix = "arXiv",
    primaryClass = "astro-ph.CO",
    doi = "10.1103/PhysRevD.88.043502",
    journal = "Phys. Rev. D",
    volume = "88",
    pages = "043502",
    year = "2013"
}

@ARTICLE{2016A&A...594A..91L,
       author = {{L{\'o}pez}, S. and {D'Odorico}, V. and {Ellison}, S.~L. and {Becker}, G.~D. and {Christensen}, L. and {Cupani}, G. and {Denney}, K.~D. and {P{\^a}ris}, I. and {Worseck}, G. and {Berg}, T.~A.~M. and {Cristiani}, S. and {Dessauges-Zavadsky}, M. and {Haehnelt}, M. and {Hamann}, F. and {Hennawi}, J. and {Ir{\v{s}}i{\v{c}}}, V. and {Kim}, T. -S. and {L{\'o}pez}, P. and {Lund Saust}, R. and {M{\'e}nard}, B. and {Perrotta}, S. and {Prochaska}, J.~X. and {S{\'a}nchez-Ram{\'\i}rez}, R. and {Vestergaard}, M. and {Viel}, M. and {Wisotzki}, L.},
        title = "{XQ-100: A legacy survey of one hundred 3.5 {\ensuremath{\lesssim}} z {\ensuremath{\lesssim}} 4.5 quasars observed with VLT/X-shooter}",
      journal = {Astron. Astrophys.},
         year = 2016,
        month = oct,
       volume = {594},
          eid = {A91},
        pages = {A91},
          doi = {10.1051/0004-6361/201628161},
archivePrefix = {arXiv},
       eprint = {1607.08776},
 primaryClass = {astro-ph.GA},
       adsurl = {https://ui.adsabs.harvard.edu/abs/2016A&A...594A..91L}
}

@article{BOSS:2013rpr,
    author = "Palanque-Delabrouille, Nathalie and others",
    collaboration = "BOSS",
    title = "{The one-dimensional Ly-alpha forest power spectrum from BOSS}",
    eprint = "1306.5896",
    archivePrefix = "arXiv",
    primaryClass = "astro-ph.CO",
    doi = "10.1051/0004-6361/201322130",
    journal = "Astron. Astrophys.",
    volume = "559",
    pages = "A85",
    year = "2013"
}

@article{Rogers:2023upm,
    author = "Rogers, Keir K. and Poulin, Vivian",
    title = "{5\ensuremath{\sigma} tension between Planck cosmic microwave background and eBOSS Lyman-alpha forest and constraints on physics beyond \ensuremath{\Lambda}CDM}",
    eprint = "2311.16377",
    archivePrefix = "arXiv",
    primaryClass = "astro-ph.CO",
    doi = "10.1103/PhysRevResearch.7.L012018",
    journal = "Phys. Rev. Res.",
    volume = "7",
    number = "1",
    pages = "L012018",
    year = "2025"
}

@article{Montefalcone:2025akm,
    author = "Montefalcone, Gabriele and Hooper, Dan and Freese, Katherine and Kelso, Chris and Kuhnel, Florian and Sandick, Pearl",
    title = "{Does Memory Burden Open a New Mass Window for Primordial Black Holes as Dark Matter?}",
    eprint = "2503.21005",
    archivePrefix = "arXiv",
    primaryClass = "astro-ph.CO",
    reportNumber = "UTWI-09-2025, NORDITA-2025-015",
    month = "3",
    year = "2025"
}

@article{Escriva:2021aeh,
    author = "Escriv{\`a}, Albert",
    title = "{PBH Formation from Spherically Symmetric Hydrodynamical Perturbations: A Review}",
    eprint = "2111.12693",
    archivePrefix = "arXiv",
    primaryClass = "gr-qc",
    doi = "10.3390/universe8020066",
    journal = "Universe",
    volume = "8",
    number = "2",
    pages = "66",
    year = "2022"
}

@article{Garcia-Gallego:2025kiw,
    author = "Garcia-Gallego, Olga and Ir{\v{s}}i{\v{c}}, Vid and Haehnelt, Martin G. and Viel, Matteo and Bolton, James S.",
    title = "{Constraining mixed dark matter models with high-redshift Lyman-alpha forest data}",
    eprint = "2504.06367",
    archivePrefix = "arXiv",
    primaryClass = "astro-ph.CO",
    doi = "10.1103/4k29-h99l",
    journal = "Phys. Rev. D",
    volume = "112",
    number = "4",
    pages = "043502",
    year = "2025"
}

@article{Irsic:2023equ,
    author = "Ir{\v{s}}i{\v{c}}, Vid and others",
    title = "{Unveiling dark matter free streaming at the smallest scales with the high redshift Lyman-alpha forest}",
    eprint = "2309.04533",
    archivePrefix = "arXiv",
    primaryClass = "astro-ph.CO",
    doi = "10.1103/PhysRevD.109.043511",
    journal = "Phys. Rev. D",
    volume = "109",
    number = "4",
    pages = "043511",
    year = "2024"
}

@article{Schneider:2016uqi,
    author = "Schneider, Aurel",
    title = "{Astrophysical constraints on resonantly produced sterile neutrino dark matter}",
    eprint = "1601.07553",
    archivePrefix = "arXiv",
    primaryClass = "astro-ph.CO",
    doi = "10.1088/1475-7516/2016/04/059",
    journal = "JCAP",
    volume = "04",
    pages = "059",
    year = "2016"
}

@article{Egana-Ugrinovic:2021gnu,
    author = "Egana-Ugrinovic, Daniel and Essig, Rouven and Gift, Daniel and LoVerde, Marilena",
    title = "{The Cosmological Evolution of Self-interacting Dark Matter}",
    eprint = "2102.06215",
    archivePrefix = "arXiv",
    primaryClass = "astro-ph.CO",
    doi = "10.1088/1475-7516/2021/05/013",
    journal = "JCAP",
    volume = "05",
    pages = "013",
    year = "2021"
}

@article{Bernal:2020kse,
    author = "Bernal, Nicol{\'a}s and Zapata, {\'O}scar",
    title = "{Self-interacting Dark Matter from Primordial Black Holes}",
    eprint = "2010.09725",
    archivePrefix = "arXiv",
    primaryClass = "hep-ph",
    reportNumber = "PI/UAN-2020-680FT",
    doi = "10.1088/1475-7516/2021/03/007",
    journal = "JCAP",
    volume = "03",
    pages = "007",
    year = "2021"
}

@article{Bernal:2020ili,
    author = "Bernal, Nicol{\'a}s and Zapata, {\'O}scar",
    title = "{Gravitational dark matter production: primordial black holes and UV freeze-in}",
    eprint = "2011.02510",
    archivePrefix = "arXiv",
    primaryClass = "hep-ph",
    reportNumber = "PI/UAN-2020-682FT",
    doi = "10.1016/j.physletb.2021.136129",
    journal = "Phys. Lett. B",
    volume = "815",
    pages = "136129",
    year = "2021"
}

@article{Chaudhuri:2020wjo,
    author = "Chaudhuri, A. and Dolgov, A.",
    title = "{PBH Evaporation, Baryon Asymmetry, and Dark Matter}",
    eprint = "2001.11219",
    archivePrefix = "arXiv",
    primaryClass = "astro-ph.CO",
    doi = "10.1134/S1063776121110078",
    journal = "J. Exp. Theor. Phys.",
    volume = "133",
    number = "5",
    pages = "552--566",
    year = "2021"
}

@article{Hooper:2019gtx,
    author = "Hooper, Dan and Krnjaic, Gordan and McDermott, Samuel D.",
    title = "{Dark Radiation and Superheavy Dark Matter from Black Hole Domination}",
    eprint = "1905.01301",
    archivePrefix = "arXiv",
    primaryClass = "hep-ph",
    reportNumber = "FERMILAB-PUB-19-186-A",
    doi = "10.1007/JHEP08(2019)001",
    journal = "JHEP",
    volume = "08",
    pages = "001",
    year = "2019"
}

@article{Allahverdi:2017sks,
    author = "Allahverdi, Rouzbeh and Dent, James and Osinski, Jacek",
    title = "{Nonthermal production of dark matter from primordial black holes}",
    eprint = "1711.10511",
    archivePrefix = "arXiv",
    primaryClass = "astro-ph.CO",
    doi = "10.1103/PhysRevD.97.055013",
    journal = "Phys. Rev. D",
    volume = "97",
    number = "5",
    pages = "055013",
    year = "2018"
}

@article{Dai:2009hx,
    author = "Dai, De-Chang and Freese, Katherine and Stojkovic, Dejan",
    title = "{Constraints on dark matter particles charged under a hidden gauge group from primordial black holes}",
    eprint = "0904.3331",
    archivePrefix = "arXiv",
    primaryClass = "hep-ph",
    doi = "10.1088/1475-7516/2009/06/023",
    journal = "JCAP",
    volume = "06",
    pages = "023",
    year = "2009"
}

@article{Khlopov:2004tn,
    author = "Khlopov, M. Yu. and Barrau, Aurelien and Grain, Julien",
    title = "{Gravitino production by primordial black hole evaporation and constraints on the inhomogeneity of the early universe}",
    eprint = "astro-ph/0406621",
    archivePrefix = "arXiv",
    doi = "10.1088/0264-9381/23/6/004",
    journal = "Class. Quant. Grav.",
    volume = "23",
    pages = "1875--1882",
    year = "2006"
}

@article{Green:1999yh,
    author = "Green, Anne M.",
    title = "{Supersymmetry and primordial black hole abundance constraints}",
    eprint = "astro-ph/9903484",
    archivePrefix = "arXiv",
    doi = "10.1103/PhysRevD.60.063516",
    journal = "Phys. Rev. D",
    volume = "60",
    pages = "063516",
    year = "1999"
}

@article{Bernal:2020bjf,
    author = "Bernal, Nicol{\'a}s and Zapata, {\'O}scar",
    title = "{Dark Matter in the Time of Primordial Black Holes}",
    eprint = "2011.12306",
    archivePrefix = "arXiv",
    primaryClass = "astro-ph.CO",
    reportNumber = "PI/UAN-2020-683FT",
    doi = "10.1088/1475-7516/2021/03/015",
    journal = "JCAP",
    volume = "03",
    pages = "015",
    year = "2021"
}

@article{Morrison:2018xla,
    author = "Morrison, Logan and Profumo, Stefano and Yu, Yan",
    title = "{Melanopogenesis: Dark Matter of (almost) any Mass and Baryonic Matter from the Evaporation of Primordial Black Holes weighing a Ton (or less)}",
    eprint = "1812.10606",
    archivePrefix = "arXiv",
    primaryClass = "astro-ph.CO",
    doi = "10.1088/1475-7516/2019/05/005",
    journal = "JCAP",
    volume = "05",
    pages = "005",
    year = "2019"
}

@article{Masina:2020xhk,
    author = "Masina, Isabella",
    title = "{Dark matter and dark radiation from evaporating primordial black holes}",
    eprint = "2004.04740",
    archivePrefix = "arXiv",
    primaryClass = "hep-ph",
    doi = "10.1140/epjp/s13360-020-00564-9",
    journal = "Eur. Phys. J. Plus",
    volume = "135",
    number = "7",
    pages = "552",
    year = "2020"
}

@article{Auffinger:2020afu,
    author = "Auffinger, J{\'e}r{\'e}my and Masina, Isabella and Orlando, Giorgio",
    title = "{Bounds on warm dark matter from Schwarzschild primordial black holes}",
    eprint = "2012.09867",
    archivePrefix = "arXiv",
    primaryClass = "hep-ph",
    doi = "10.1140/epjp/s13360-021-01247-9",
    journal = "Eur. Phys. J. Plus",
    volume = "136",
    number = "2",
    pages = "261",
    year = "2021"
}

@article{Gondolo:2020uqv,
    author = "Gondolo, Paolo and Sandick, Pearl and Shams Es Haghi, Barmak",
    title = "{Effects of primordial black holes on dark matter models}",
    eprint = "2009.02424",
    archivePrefix = "arXiv",
    primaryClass = "hep-ph",
    doi = "10.1103/PhysRevD.102.095018",
    journal = "Phys. Rev. D",
    volume = "102",
    number = "9",
    pages = "095018",
    year = "2020"
}

@article{DEramo:2020gpr,
    author = "D'Eramo, Francesco and Lenoci, Alessandro",
    title = "{Lower mass bounds on FIMP dark matter produced via freeze-in}",
    eprint = "2012.01446",
    archivePrefix = "arXiv",
    primaryClass = "hep-ph",
    reportNumber = "DESY 20-219, DESY-20-219",
    doi = "10.1088/1475-7516/2021/10/045",
    journal = "JCAP",
    volume = "10",
    pages = "045",
    year = "2021"
}

@article{Kim:2025kgu,
    author = "Kim, TaeHun and Gong, Jinn-Ouk and Jeong, Donghui and Jung, Dong-Won and Kim, Yeong Gyun and Lee, Kang Young",
    title = "{Planck isocurvature constraint on primordial black holes lighter than a kiloton}",
    eprint = "2503.14581",
    archivePrefix = "arXiv",
    primaryClass = "astro-ph.CO",
    reportNumber = "APCTP-Pre2025-005",
    month = "3",
    year = "2025"
}

@article{Samanta:2021mdm,
    author = "Samanta, Rome and Urban, Federico R.",
    title = "{Testing super heavy dark matter from primordial black holes with gravitational waves}",
    eprint = "2112.04836",
    archivePrefix = "arXiv",
    primaryClass = "hep-ph",
    doi = "10.1088/1475-7516/2022/06/017",
    journal = "JCAP",
    volume = "06",
    number = "06",
    pages = "017",
    year = "2022"
}

@article{Bernal:2021bbv,
    author = "Bernal, Nicol{\'a}s and Perez-Gonzalez, Yuber F. and Xu, Yong and Zapata, {\'O}scar",
    title = "{ALP dark matter in a primordial black hole dominated universe}",
    eprint = "2110.04312",
    archivePrefix = "arXiv",
    primaryClass = "hep-ph",
    reportNumber = "PI/UAN-2021-702FT, FERMILAB-PUB-21-478-T, NUHEP-TH/21-16, IPPP/21/36",
    doi = "10.1103/PhysRevD.104.123536",
    journal = "Phys. Rev. D",
    volume = "104",
    number = "12",
    pages = "123536",
    year = "2021"
}

@article{Cheek:2021cfe,
    author = "Cheek, Andrew and Heurtier, Lucien and Perez-Gonzalez, Yuber F. and Turner, Jessica",
    title = "{Primordial black hole evaporation and dark matter production. II. Interplay with the freeze-in or freeze-out mechanism}",
    eprint = "2107.00016",
    archivePrefix = "arXiv",
    primaryClass = "hep-ph",
    reportNumber = "FERMILAB-PUB-21-305-T, NUHEP-TH/21-07, CP3-21-42, IPPP/21/01",
    doi = "10.1103/PhysRevD.105.015023",
    journal = "Phys. Rev. D",
    volume = "105",
    number = "1",
    pages = "015023",
    year = "2022"
}

@article{Cheek:2021odj,
    author = "Cheek, Andrew and Heurtier, Lucien and Perez-Gonzalez, Yuber F. and Turner, Jessica",
    title = "{Primordial black hole evaporation and dark matter production. I. Solely Hawking radiation}",
    eprint = "2107.00013",
    archivePrefix = "arXiv",
    primaryClass = "hep-ph",
    reportNumber = "FERMILAB-PUB-21-304-T, NUHEP-TH/21-06, CP3-21-41, IPPP/21/02",
    doi = "10.1103/PhysRevD.105.015022",
    journal = "Phys. Rev. D",
    volume = "105",
    number = "1",
    pages = "015022",
    year = "2022"
}

@article{Barman:2021ost,
    author = "Barman, Basabendu and Borah, Debasish and Das, Suruj Jyoti and Roshan, Rishav",
    title = "{Non-thermal origin of asymmetric dark matter from inflaton and primordial black holes}",
    eprint = "2111.08034",
    archivePrefix = "arXiv",
    primaryClass = "hep-ph",
    reportNumber = "PI/UAN-2021-705FT",
    doi = "10.1088/1475-7516/2022/03/031",
    journal = "JCAP",
    volume = "03",
    number = "03",
    pages = "031",
    year = "2022"
}

@article{Chen:2023tzd,
    author = "Chen, Muping and Gelmini, Graciela B. and Lu, Philip and Takhistov, Volodymyr",
    title = "{Primordial black hole sterile neutrinogenesis: sterile~neutrino dark matter production independent of couplings}",
    eprint = "2312.12136",
    archivePrefix = "arXiv",
    primaryClass = "hep-ph",
    reportNumber = "KEK-QUP-2023-0035, KEK-TH-2584, KEK-Cosmo-0334, IPMU23-0049",
    doi = "10.1088/1475-7516/2024/07/059",
    journal = "JCAP",
    volume = "07",
    pages = "059",
    year = "2024"
}

@article{Fujita:2014hha,
    author = "Fujita, Tomohiro and Kawasaki, Masahiro and Harigaya, Keisuke and Matsuda, Ryo",
    title = "{Baryon asymmetry, dark matter, and density perturbation from primordial black holes}",
    eprint = "1401.1909",
    archivePrefix = "arXiv",
    primaryClass = "astro-ph.CO",
    reportNumber = "IPMU-14-0009, ICRR-REPORT-668-2013-17",
    doi = "10.1103/PhysRevD.89.103501",
    journal = "Phys. Rev. D",
    volume = "89",
    number = "10",
    pages = "103501",
    year = "2014"
}

@article{Bell:1998jk,
    author = "Bell, Nicole F. and Volkas, R. R.",
    title = "{Mirror matter and primordial black holes}",
    eprint = "astro-ph/9812301",
    archivePrefix = "arXiv",
    reportNumber = "UM-P-98-62, RCHEP-98-18",
    doi = "10.1103/PhysRevD.59.107301",
    journal = "Phys. Rev. D",
    volume = "59",
    pages = "107301",
    year = "1999"
}

@inproceedings{Matsas:1998zm,
    author = "Matsas, G. E. A. and Montero, J. C. and Pleitez, V. and Vanzella, D. A. T.",
    title = "{Dark matter: The Top of the iceberg?}",
    booktitle = "{Conference on Topics in Theoretical Physics II: Festschrift for A.H. Zimerman}",
    eprint = "hep-ph/9810456",
    archivePrefix = "arXiv",
    reportNumber = "IFT-P-075-98",
    month = "10",
    year = "1998"
}

@article{Baumann:2007yr,
    author = "Baumann, Daniel and Steinhardt, Paul J. and Turok, Neil",
    title = "{Primordial Black Hole Baryogenesis}",
    eprint = "hep-th/0703250",
    archivePrefix = "arXiv",
    reportNumber = "PUPT-2229",
    month = "3",
    year = "2007"
}

@article{Dolgov:2000ht,
    author = "Dolgov, A. D. and Naselsky, P. D. and Novikov, I. D.",
    title = "{Gravitational waves, baryogenesis, and dark matter from primordial black holes}",
    eprint = "astro-ph/0009407",
    archivePrefix = "arXiv",
    month = "9",
    year = "2000"
}

@article{Euclid:2024pwi,
    author = "Lesgourgues, J. and others",
    collaboration = "Euclid",
    title = "{Euclid preparation - LVI. Sensitivity to non-standard particle dark matter models}",
    eprint = "2406.18274",
    archivePrefix = "arXiv",
    primaryClass = "astro-ph.CO",
    reportNumber = "TTK-24-26",
    doi = "10.1051/0004-6361/202451611",
    journal = "Astron. Astrophys.",
    volume = "693",
    pages = "A249",
    year = "2025"
}

@article{Hooper:2022byl,
    author = {Hooper, Deanna C. and Sch{\"o}neberg, Nils and Murgia, Riccardo and Archidiacono, Maria and Lesgourgues, Julien and Viel, Matteo},
    title = "{One likelihood to bind them all: Lyman-{\ensuremath{\alpha}} constraints on non-standard dark matter}",
    eprint = "2206.08188",
    archivePrefix = "arXiv",
    primaryClass = "astro-ph.CO",
    reportNumber = "HIP-2022-17/TH, TTK-22-21",
    doi = "10.1088/1475-7516/2022/10/032",
    journal = "JCAP",
    volume = "10",
    pages = "032",
    year = "2022"
}

@article{Viel:2005qj,
    author = "Viel, Matteo and Lesgourgues, Julien and Haehnelt, Martin G. and Matarrese, Sabino and Riotto, Antonio",
    title = "{Constraining warm dark matter candidates including sterile neutrinos and light gravitinos with WMAP and the Lyman-alpha forest}",
    eprint = "astro-ph/0501562",
    archivePrefix = "arXiv",
    reportNumber = "DFPD-05-08, LAPTH-1086-05",
    doi = "10.1103/PhysRevD.71.063534",
    journal = "Phys. Rev. D",
    volume = "71",
    pages = "063534",
    year = "2005"
}

@article{Wang:2025pum,
    author = "Wang, Tianning and Grohs, Evan and Mersini-Houghton, Laura",
    title = "{How Primordial Black Holes Change BBN}",
    eprint = "2511.18646",
    archivePrefix = "arXiv",
    primaryClass = "astro-ph.CO",
    month = "11",
    year = "2025"
}

@article{Kitabayashi:2025iaq,
    author = "Kitabayashi, Teruyuki and Takeshita, Amane",
    title = "{WIMP/FIMP dark matter and primordial black holes with memory burden effect}",
    eprint = "2506.20071",
    archivePrefix = "arXiv",
    primaryClass = "hep-ph",
    month = "6",
    year = "2025"
}

@article{Dvali:2025ktz,
    author = "Dvali, Gia and Zantedeschi, Michael and Zell, Sebastian",
    title = "{Transitioning to Memory Burden: Detectable Small Primordial Black Holes as Dark Matter}",
    eprint = "2503.21740",
    archivePrefix = "arXiv",
    primaryClass = "hep-ph",
    month = "3",
    year = "2025"
}

@article{Borah:2024bcr,
    author = "Borah, Debasish and Das, Nayan",
    title = "{Successful cogenesis of baryon and dark matter from memory-burdened PBH}",
    eprint = "2410.16403",
    archivePrefix = "arXiv",
    primaryClass = "hep-ph",
    doi = "10.1088/1475-7516/2025/02/031",
    journal = "JCAP",
    volume = "02",
    pages = "031",
    year = "2025"
}

@article{Shallue:2024hqe,
    author = "Shallue, Christopher J. and Mu{\~n}oz, Julian B. and Krnjaic, Gordan Z.",
    title = "{Warm Hawking relics from primordial black hole domination}",
    eprint = "2406.08535",
    archivePrefix = "arXiv",
    primaryClass = "astro-ph.CO",
    reportNumber = "FERMILAB-PUB-24-0298-T",
    doi = "10.1088/1475-7516/2025/02/026",
    journal = "JCAP",
    volume = "02",
    pages = "026",
    year = "2025"
}

@article{Bertuzzo:2024fns,
    author = "Bertuzzo, Enrico and Perez-Gonzalez, Yuber F. and Salla, Gabriel M. and Funchal, Renata Zukanovich",
    title = "{Gravitationally produced dark matter and primordial black holes}",
    eprint = "2405.17611",
    archivePrefix = "arXiv",
    primaryClass = "hep-ph",
    reportNumber = "IPPP/24/25",
    doi = "10.1088/1475-7516/2024/09/059",
    journal = "JCAP",
    volume = "09",
    pages = "059",
    year = "2024"
}

\end{document}